\documentclass[sigconf, nonacm]{acmart}

\usepackage{amsmath,amssymb,amsfonts}
\usepackage{graphicx}
\usepackage{textcomp}
\usepackage{bmpsize}
\usepackage{comment}
\usepackage{enumitem}
\usepackage{amssymb}% http://ctan.org/pkg/amssymb
\usepackage{pifont}% http://ctan.org/pkg/pifont
\newcommand{\cmark}{\ding{51}}%
\newcommand{\xmark}{\ding{55}}%
\usepackage{makecell}

\usepackage{tikz}
 \UseRawInputEncoding
\newtheorem{definition}{Definition}
\def\Plus{\texttt{+}}

\usepackage{xcolor}

\newcommand{\gmnote}[1]{\textcolor{red}{*Gonzalo: #1}}

\setcitestyle{numbers}

\usepackage{amsthm}% for \@addpunct
\makeatletter
\def\els@aparagraph[#1]#2{\elsparagraph[#1]{#2\@addpunct{.}}}
\def\els@bparagraph#1{\elsparagraph*{#1\@addpunct{.}}}
\makeatother

\usepackage{caption} 
\usepackage{float}
\floatstyle{plaintop}
\restylefloat{table}

\usepackage{xcolor}

\usepackage{array}
\newcolumntype{M}[1]{>{\centering\arraybackslash}m{#1}}

\usepackage{afterpage}

\usepackage{comment}

\usepackage{booktabs}

\def\Plus{\texttt{+}}

\usepackage{pdfpages}

\usepackage{algorithmicx}
\usepackage[ruled, vlined]{algorithm2e}
\usepackage{float}

\usepackage{scalerel,stackengine}
\stackMath
\newcommand\reallywidehat[1]{%
\savestack{\tmpbox}{\stretchto{%
  \scaleto{%
    \scalerel*[\widthof{\ensuremath{#1}}]{\kern-.6pt\bigwedge\kern-.6pt}%
    {\rule[-\textheight/2]{1ex}{\textheight}}%WIDTH-LIMITED BIG WEDGE
  }{\textheight}% 
}{0.5ex}}%
\stackon[1pt]{#1}{\tmpbox}%
}

\usepackage{color}

\setcopyright{none}
\copyrightyear{2023}
\acmYear{2023}
\acmDOI{XXXXXXX.XXXXXXX}
\acmISBN{}
\acmConference{}{}{}
\renewcommand\footnotetextcopyrightpermission[1]{}
\begin{document}

\title{Lessons Learned: Surveying the Practicality of Differential Privacy in the Industry}

\author{Gonzalo Munilla Garrido}
\email{gonzalo.munilla-garrido@tum.de}
\affiliation{
\institution{TUM}
\country{Germany}
}

\author{Xiaoyuan Liu}
\email{xiaoyuanliu@berkeley.edu }
\affiliation{
  \institution{UC Berkeley}
  \country{USA}
}

\author{Florian Matthes}
\email{matthes@tum.de}
\affiliation{
  \institution{TUM}
  \country{Germany}
}

\author{Dawn Song}
\email{dawnsong@berkeley.edu}
\affiliation{
  \institution{UC Berkeley}
  \country{USA}
}

\begin{abstract}
Since its introduction in 2006, differential privacy has emerged as a predominant statistical tool for quantifying data privacy in academic works. Yet despite the plethora of research and open-source utilities that have accompanied its rise, with limited exceptions, differential privacy has failed to achieve widespread adoption in the enterprise domain. Our study aims to shed light on the fundamental causes underlying this academic-industrial utilization gap through detailed interviews of 24 privacy practitioners across 9 major companies. We analyze the results of our survey to provide key findings and suggestions for companies striving to improve privacy protection in their data workflows and highlight the necessary and missing requirements of existing differential privacy tools, with the goal of guiding researchers working towards the broader adoption of differential privacy. Our findings indicate that analysts suffer from lengthy bureaucratic processes for requesting access to sensitive data, yet once granted, only scarcely-enforced privacy policies stand between rogue practitioners and misuse of private information. We thus argue that differential privacy can significantly improve the processes of requesting and conducting data exploration across silos, and conclude that with a few of the improvements suggested herein, the practical use of differential privacy across the enterprise is within striking distance.
\end{abstract}

\keywords{Privacy-enhancing technology, data sharing, data analytics, platform, SQL, machine learning, case study, interviews, survey}

\maketitle

\section{Introduction}
\label{sec:Introduction}

Several factors have spurred the development of more advanced privacy-enhancing technologies (PETs) in the past years.
On the one hand, from an adversarial perspective, (i) multiple white-hat \emph{attacks} have shown that ``traditional'' anonymization techniques such as suppressing names are vulnerable to re-identification across industries~\cite{narayanan_robust_2008, gao_elastic_2014, kondor_towards_2020, dwork_exposed_2017, sweeney_identifying_2013, archie_anonymization_nodate}.
Additionally, between 2020 and 2021, (ii) the total cost of \emph{data breaches} have increased by $10\%$ on average~\cite{IBM_data_breach}. 
Moreover, (iii) governments have promulgated \emph{data protection laws} in the past years, such as the European General Data Protection Regulation (GDPR)~\cite{GDPR_EU2016} or the California Consumer Privacy Act~\cite{ccpa2018}. 
In particular, the GDPR has issued fines as high as \$$887$M~\cite{amazon-fine} and \$$120$M~\cite{amazon-google-fine}.
Furthermore, (iv) beyond the \emph{ethical and moral obligations} of companies to protect people's privacy, providing the best privacy protection available could (v) \emph{differentiate and appreciate their brands}~\cite{data_exchange_McKinsey}, (vi) provide \emph{fairer products and services} that avoid price discrimination~\cite{privacy_economic_good}, and (vii) \emph{increase data collection} as PETs help to surmount regulatory barriers fairly~\cite{kaaniche2016abs}.
Aiming to materialize these benefits while mitigating the privacy risks, researchers have turned to differential privacy (DP), which, since its inception in 2006 by Dwork et al.~\cite{DP_original}, has become the golden privacy standard in academia due to its unique privacy guarantees. 
%, e.g., the average cost of a breach in healthcare has risen from \$$7.1$M to \$$9.2$M~\cite{IBM_data_breach}.

However, despite numerous open-source utilities, only a few tech companies~\cite{Apple_DP, Google_apple_local_dp, microsoft_DP} and the US Census Bureau~\cite{census_DP} have adopted DP.
Accordingly, our work addresses the research gap in bringing DP into organizations' workflows and reaching broader adoption. Dwork~et~al.~\cite{dwork_differential_2019} partly covered the gap by interviewing DP experts, while our study closes the remaining gap by bringing non-experts into the spotlight.
Thus, we interviewed $24$ practitioners ($19$ analysts and $5$ data stewards) across $9$ major companies that have not yet deployed DP.
Overall, our main contributions are: 
\begin{enumerate}[label=(\roman*)]
    \item \textbf{Survey results.} We formulated $5$ research questions and derived $24$ interview questions thereof. The results of the interviews provide an overview of the current state of data access models (\S~\ref{subsec:Accessing_Data_in_the_Industry}), privacy practices (\S~\ref{subsec:Anonymizing_Data}), motivation behind privacy protection (\S~\ref{subsec:Motivating_Privacy}), and analysis workflows (\S~\ref{subsec:The_Feasibility_of_Differential_Privacy}) in the industry. 
    \item \textbf{Key findings}. From the survey results, we extract $11$ key findings, suggest improvements, and answer the $5$ research questions about the practicality of DP in the industry (\S~\ref{sec:Key_Findings}).
    \item \textbf{Functional requirements}. Based on the key findings, we propose $10$ key desiderata to guide organizations in building privacy-enhancing analytics systems that tackle the privacy-related pain points in their workflows (\S~\ref{subsec:desiderata}).
    \item \textbf{Missing building blocks}. Given the identified key desiderata, we outline $7$ gaps in state-of-the-art DP tooling (\S~\ref{subsec:gap_in_DP}). 
\end{enumerate}

\textit{Privacy officers} and \textit{legal practitioners} will find (i) and (ii) helpful in understanding the landscape of privacy and analysis workflows in the industry.
\textit{Software engineers} and \textit{developers} will also appreciate (iii) and (iv) as these contributions focus on tooling, and, additionally, will find our early-stage privacy-enhancing analytics system design presented in Appendix~\ref{app_sec:system_design} helpful.
Overall, notable findings reveal that cumbersome data request processes block analysts for significant periods for every new project. Additionally, we note that SQL was more important than machine learning, and data stewards are more concerned with security than privacy. 
%We also identified the malpractice of deferring the team's responsibility of requesting and managing data to a single individual.
We conclude that DP could shorten data access processes, enable data exploration across silos, and is applicable to specific use cases. 
Moreover, DP tool designers can learn from one another as no tool outperforms the rest in every aspect, and, most importantly, bridging the gap between theory and practice is primarily an engineering problem within striking distance.
% We observed that companies are still far from having ``\textit{data at their fingertips}'', and data quality is still a predominant issue that impedes systems to completely block analysts from ``seeing'' the data, because for debugging they often need to visually inspect records.
% Overall, we observed that the major problems are (i) the lengthy data access processes nd often the protection relies on policy once access is granted.

\begin{comment}

, e.g., e-commerce~\cite{archie_anonymization_nodate}, health care~\cite{sweeney_identifying_2013}, entertainment~\cite{narayanan_robust_2008}, automotive, and telecommunications~\cite{gao_elastic_2014, kondor_towards_2020}, among others~\cite{dwork_exposed_2017}.
\textbf{Our goal is to understand the gap between theory and wide adoption of differential privacy in practice}

\end{comment}

\section{Differential Privacy}
\label{sec:differential_Privacy}

Unlike traditional privacy techniques, which are vulnerable to auxiliary information attacks~\cite{dwork_exposed_2017, sweeney_identifying_2013, gao_elastic_2014, kondor_towards_2020, narayanan_robust_2008, archie_anonymization_nodate}, differential privacy~\cite{DP_original} mathematically formalizes a privacy guarantee agnostic to background information.  
A function guarantees differential privacy (e.g., an analytics query or a machine learning (ML) model) if it bounds the information gain that an attacker can expect from its outputs.
Aligned with this adversarial perspective, for the context of this study, we define \emph{privacy} as the prevention of an individual's re-identification~\cite{Wu2012}.

In practice, the outputs of a differentially private function are similarly likely, regardless of an individual's contribution to the input data.
This similarity is bounded by the parameter $\varepsilon$, which is inversely proportional to the strength of the privacy protection.
A randomized function $\mathcal{M}(\cdot)$ satisfies differential privacy by adding calibrated random noise, typically to a deterministic function's output. 
Formally, differential privacy is defined as~\cite{goos_lecture_nodate_2}:

\vspace{-1mm}
\begin{definition}
\label{def:DP}
  ($\varepsilon$-Differential Privacy). A randomized function $\mathcal{M}(\cdot)$ is $\varepsilon$-differentially private iff for any two datasets $D$ and $D'$ differing on at most one element, and any set of possible outputs $\mathcal{S} \subseteq
  Range(\mathcal{M})$:
    \begin{center}
      	$\mathrm{Pr}[\mathcal{M}$ $(D)$  $\in$  $\mathcal{S}]$ $\leq$ $e^\varepsilon$ $\times$ $\mathrm{Pr}[\mathcal{M}$ $(D')$  $\in$  $\mathcal{S}]$.
    \end{center}
\end{definition}
\vspace{-1mm}
\noindent We introduce other concepts useful in the context of this paper:
\noindent \textbf{Sensitivity}. Beyond $\varepsilon$, the other parameter that affects the scale of the noise is the sensitivity of the deterministic function, which determines the maximum difference of the function's outputs over all possible neighboring datasets $D$ and $D'$. \newline
\noindent \textbf{Central/Local model}. An application can add differentially private noise in the \emph{central} model after aggregating data points from different clients or in the \emph{local} model by adding noise to each data point individually. While the local model requires less trust assumptions with the aggregator, it is usually noisier than the central model. \newline
\noindent  \textbf{Sequential composition}. Differential privacy algorithms follow \textit{sequential composition}~\cite{goos_lecture_nodate_2}, i.e., if one executes a sequence of (possibly different) DP mechanisms $n$ times over $D$ with $\varepsilon_i$, the consumed \emph{privacy budget} $\varepsilon = \sum \varepsilon_i$. \newline
\noindent \textbf{Privacy budget tracker}. Because the added noise is centered around $0$, an attacker could reverse engineer the $n$ outputs by averaging out the noise. Thus, systems should implement privacy budget trackers to prevent this attack. \newline
\noindent \textbf{Floating-point vulnerability}. Proofs of differential privacy mechanisms work on continuous distributions, which leads to privacy bugs in practice as the implementations rely on floating-point arithmetic~\cite{mironov_significance_2012}. There are a few solutions to this problem. In short, Mironov's Snapping mechanism~\cite{mironov_significance_2012} discards the least-significant bit in a post-processing step, Naoise~et.~al~\cite{random_sampling_DP} combine four random samples, and Haney~et.~al~\cite{precision_attacks} designed a variant of the Laplace mechanism that avoids a precision-based attack.

\section{Related Work}
\label{sec:Related_Work}

Some organizations have developed and deployed differential privacy tooling and have documented their purpose.
Specifically, Apple~\cite{Apple_DP, Google_apple_local_dp}, Google~\cite{Google_apple_local_dp}, and Microsoft~\cite{microsoft_DP} employ algorithms based on the local model of differential privacy to collect information from users.
The local model is not as predominant in the industry as the global model (our focus), which has seen more deployments in the past years:
Google's Plume~\cite{Google_Plum} enables simple statistics (count, mean, sum, variancer, and quantile) over large-scale datasets. Moreover, LinkedIn~\cite{Linkedin_DP, Linkedin_usenix, rogers_linkedins_2021} proposed an API to analyse user data, and the U.S. Census Bureau in 2020~\cite{census_DP} released microdata; however, these two approaches only considered count queries.
Additionally, there exist open-source differential privacy \emph{libraries}, \emph{frameworks}, and \emph{systems} from Google~\cite{Google_repo, priv_on_beam, ZetaSQL, zetaSQL_paper, tensor_flow_priv}, Harvard~\cite{SmartNoise_repo, gaboardi_programming_nodate}, IBM~\cite{IBM_repo}, Meta~\cite{Opacus}, OpenMined~\cite{pipelineDP} (experimental product), Tumult Labs~\cite{Tumult_Analytics}, and the University of Pennsylvania~\cite{narayan_djoin_nodate} and Texas~\cite{roy_airavat_nodate}.
Note that OpenDP encapsulates SmartNoise core~\cite{opendp_smartnoise}.
Additionally, researchers have also developed open-source systems focused on \emph{user interfaces} for differentially private analytics: Bittner~et.~al~\cite{bittner}, DPcomp~\cite{DPcomp}, DPP~\cite{DPP}, Overlook~\cite{overlook_DP}, PSI~($\Psi$)~\cite{psi}, and ViP~\cite{nanayakkara_visualizing_2022}.
However, only a few libraries have been discussed in a utility benchmark~\cite{DP_benchmark}.
Moreover, Johnson et al.'s work on differentially private SQL~\cite{Chorus_repo} at Uber~\cite{Uber_DP}
focused on a quantitative evaluation of the queries without discussing its practicality with practitioners.
Unlike the previous literature above, we aim to qualitatively understand the practicality and adaptability of differential privacy in the central model to existing data analysis pipelines within an organization beyond count queries.

Among top searches of surveys related to differential privacy in digital libraries such as IEEE~\cite{IEEE_lib}, ACM~\cite{ACM}, ScienceDirect~\cite{Science_direct}, or ArXiv~\cite{arxiv}, one may notably find surveys of applications or analysis models for differential privacy in the context of social networks~\cite{DP_social_networks}, cyber physical systems such as IoT~\cite{cps_DP}, statistical learning~\cite{statistical_learning}, location-based services~\cite{location_DP_survey}, a user survey about privacy in data sharing~\cite{user_study_data_sharing}, and lessons learned from employing differential privacy in the US Census~\cite{census_issues_DP}.
Notably, Kifer~et.~al~\cite{meta_DP} distills a set of best practices and implementation details from their experience designing differential privacy systems at Meta, which we consider in our key system desiderata proposal (see section~\ref{subsec:desiderata}).
However, our work instead explores systems from companies unfamiliar with differential privacy and focuses on answering whether differential privacy could help data analysts in the broader industry.
Lastly, the closest work to ours is from Dwork~et~al.~\cite{dwork_differential_2019}. They interviewed differential privacy experts regarding their implementation specifications.
We differentiate from Dwork~et~al.~\cite{dwork_differential_2019} in that the hereby interviewed practitioners and the organizations as a whole had no significant technical expertise on differential privacy, which are the vast majority in any industry, and, specifically, we sought to understand whether differential privacy could lift the privacy-related roadblocks in their data analysis workflow.

%In addition to the interviews, and based on the distilled insights thereof, we also define critical system desiderata for a privacy-enhancing solution for the data analysis pipeline of organizations looking to improve their privacy protection in their analysis workflows.

%\vspace{-1mm}
\section{Research Method}
\label{sec:Research_Method}

While a few organizations have successfully deployed differential privacy for data analysis~\cite{Apple_DP, Google_apple_local_dp, microsoft_DP, Linkedin_DP, census_DP}, the large majority have not.
To understand whether differential privacy in the central model is practical in their analysis workflow, following a method inspired by Dwork~et~al.~\cite{dwork_differential_2019}, we performed an empirical study of a set of institutions that have not deployed differential privacy yet for their internal analysis workflows in production.
Since the focus is learning whether institutions could benefit from differential privacy, the unit of analysis is the institutions themselves.

Our study captures the answers of $24$ practitioners from $9$ organizations ($19$ data analysts/engineers and $5$ data stewards).
These organizations belong to different industries and are of different sizes (see details in Table~\ref{tab:interviewed_companies} of Appendix~\ref{app_sec:interviewed_companies}).
The jurisdictions under which the companies operate contextualize our key findings to the EU ($5$ companies) and the USA ($4$ companies).
In some organizations, we interviewed multiple practitioners to produce a holistic picture of their data analysis ecosystem.
Most interviewees held the title of \emph{data analyst}, while a few were data engineers or team leaders.
Irrespective of their title, all practitioners had at least two years and at most $10$ years of experience in the field (around $5$ years on average) and a comprehensive knowledge of their organization's tools and workflows for data analysis. \smallskip

\noindent \textbf{Interview Format and Research Questions.}
We interviewed each of the $19$ data analysts for approximately one hour through a video conference, except for three via email correspondence, between November 2021 and August 2022.
We produced the research questions (RQs) and the questionnaire prior to the interviews and based on the authors' knowledge of differential privacy and feedback from practitioners other than the ones interviewed.
The research questions aimed to understand whether differential privacy could enhance their corresponding institutions' analysis workflow by identifying missing opportunities, assessing the impact of differential privacy in their workflow, and identifying roadblocks. 

We carefully formulated the questions broadly to enable interviewees to express their views freely, recount their experiences fully, and reduce response bias and priming.
Because the organizations have not deployed differential privacy, most interviewees were not familiar with differential privacy; only two had some non-technical familiarity.
We tackled this challenge by explaining differential privacy at a high level before starting the questionnaire.
We produced the questionnaire for data analysts in Appendix~\ref{app_sec:Interview_questionnaire_analysts}, whose results are collected in section~\ref{sec:Results_of_the_Case_Studies}.
Only $4$ of the $24$ questions contained the word ``differential privacy'', which the interviewees could nonetheless answer without a deeper technical understanding (see Appendix~\ref{app_sec:Interview_questionnaire_analysts}).

Furthermore, we performed a deep dive in one corporation by interviewing $10$ analysts.
Additionally, to understand the process and motivation behind this corporation's privacy protection, we interviewed five \emph{data stewards} via video conference or email correspondence with a second questionnaire (see Appendix~\ref{app_sec:Interview_questionnaire_stewards}).
Data stewards control access to and minimize the risk of data interactions, e.g., auditing analysts' purposes before granting them access.
%and customizing datasets with the minimum data necessary to perform the analysis.
Altogether, we distill key findings and answer these $5$ RQs: \smallskip

\noindent \textbf{RQ1:} \textit{What is the context of privacy protection in the targeted organization?} 
The data stewards provided a perspective of their data protection practices, shedding light on their motivation, concerns, and possible improvements of their methods in privacy protection. \smallskip 

\noindent \textbf{RQ2:} \textit{Could differential privacy tackle the privacy-related pain points of an analysis workflow in an organization?}
The answer draws a picture of the workflow and the improvements analysts would welcome.
This holistic picture helps us identify opportunities for differential privacy in organizations' analytics workflows.
\smallskip

\noindent \textbf{RQ3:} \textit{When does differential privacy impede an analysis?}
Differential privacy is not a silver bullet; thus, we aim to explore the limitations of differential privacy in an organization.
Moreover, as the mechanisms to make SQL-like queries fulfill differential privacy are well-understood~\cite{Uber_DP, Chorus_repo, ZetaSQL}, we investigate whether this type of query is common in analysts' workflows and bring significant benefits in exchange for moderate effort. 
\smallskip

\noindent \textbf{RQ4:} \textit{How would differential privacy affect the workflow of an analyst?}
Analysts are not accustomed to the noisy outputs of differentially private mechanisms.
With this RQ, we aim to understand the impact of noise in their analysis and explore their views on different uses of differential privacy. \smallskip

% \indent \textbf{RQ5:} \textit{Is the use of SQL-like queries meaningful for their work? }
\noindent \textbf{RQ5:} \textit{Can differential privacy be applied to the frequent SQL-like~queries analysts execute?}~To exclude the impossibility of using differential privacy, we must assess whether analysts can use it in their queries.

% Finally, based on the results of this empirical study, the feedback from system engineers at the deep-dive organization, and the authors' knowledge of differential privacy and other privacy-enhancing technologies and design patterns, we specify the key desiderata for privacy-enhancing analytics systems in the context of this study (see \S~\ref{sec:Query_Re_Writer}).

\section{Results of the Privacy Study}
\label{sec:Results_of_the_Case_Studies}

To frame the research questions in the appropriate context, we first depict how the interviewed organizations usually access data and present the state-of-the-art anonymization techniques in the industry.
Subsequently, to provide a perspective on the motivation behind privacy protection, we summarize the results of the interviews with the data stewards from the deep-dive organization (RQ1).
Finally, we delve into the data analysts' answers to assess whether deploying differential privacy is useful and possible (RQ2-6).
 
 \subsection{Data Access Models}
\label{subsec:Accessing_Data_in_the_Industry}

\begin{figure}[!htbp]
    \centering
    \includegraphics[scale=0.45]{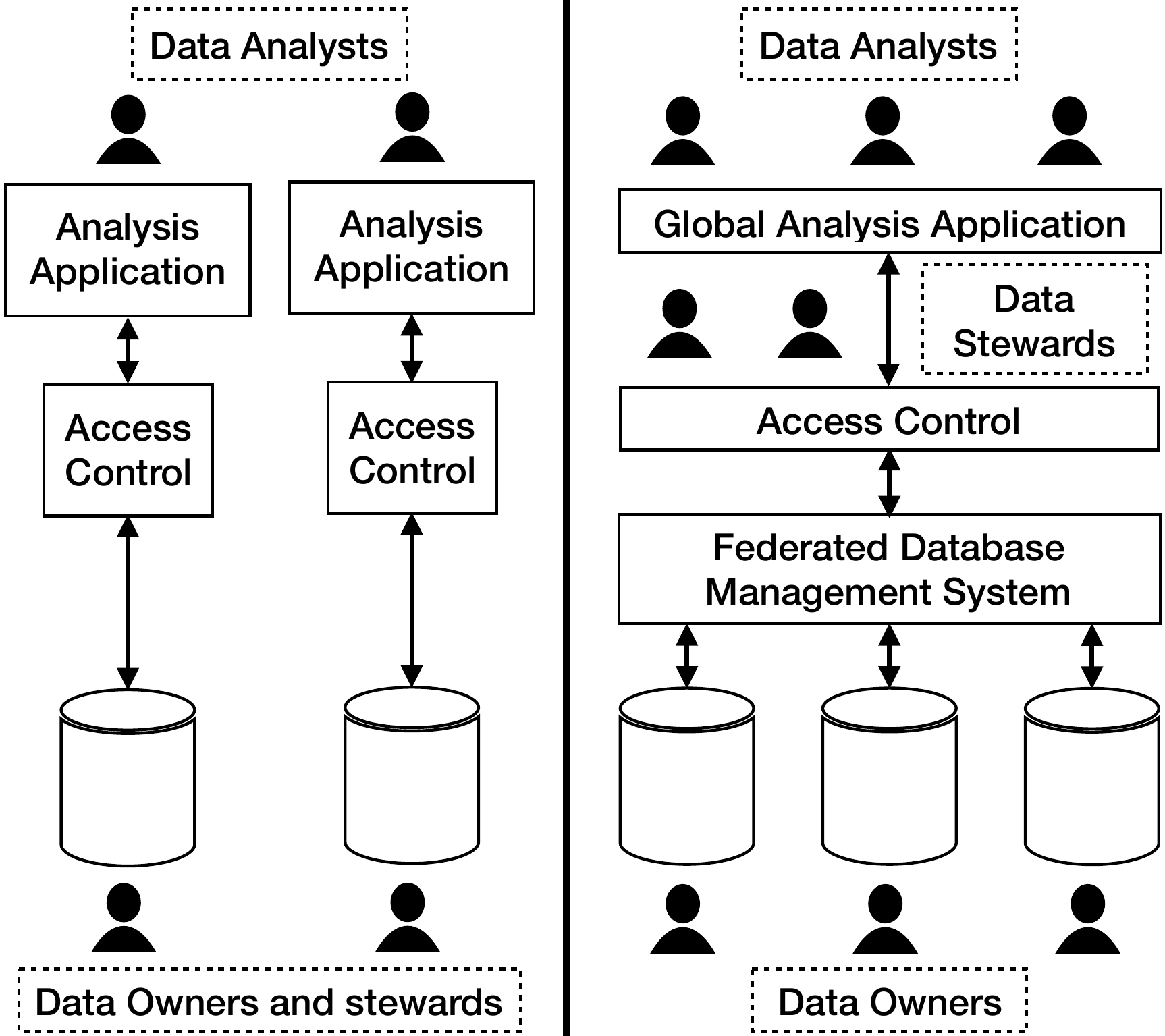}
    \caption{Informal diagram depicting a segregated (left) vs. a federated (right) models for accessing data.}
    \label{fig:segregated_federated_model}
%\vspace{-8pt}
\end{figure}

This study focuses on practitioners performing data analysis internally, i.e., without publicly releasing the results.
The interviewed organizations used one of two models for accessing data internally: \emph{segregated} and \emph{federated}~\cite{architecture_of_privacy}.
Fig.~\ref{fig:segregated_federated_model} provides an informal diagram for a quick intuition of the models.
These models used distinct roles: \emph{data owners} in charge of collecting data, \emph{data engineers} building pipelines, \emph{data stewards} assigned to overseeing the data access request processes, and \emph{data analysts} fulfilling analytics use cases.
%Analysts would only interact with the corresponding data steward when they requested dataset access.
%\vspace{-7pt}

Analytics teams in the segregated model engaged directly with data owners, whose data are stored in different data centers and regions running different systems.
The data owners would provide the data and also act as stewards.
Without an established system for automated data exchange and preparation, the analytics teams had data engineers to prepare data for every use case.
An improvement over the segregated model is its federation. 
After collection from multiple sources and pre-processing and anonymization, in the federated model, data from all domains (e.g., demographics, financial, health, etc.) are stored and easily accessible from a single application interface.
The data engineers build such data pipeline and are not usually part of an analytics team.
Data stewards guard multiple data sources, interact with analysts, and are detached from the data owner role, which is dedicated exclusively to data collection.

In both models, \emph{data protection officers} from the legal department could interact in the dataset request process, namely when analysts requested data for the first time or data were highly sensitive.

While the initial monetary investment to build a federated system could be larger than for the segregated model, the federated model holds some advantages: it (i) curtails overhead by eliminating the repetition of some tasks in the dataset access request process (e.g., user identification or analysis' purpose specification) and (ii) reduces time-intensive and cumbersome dataset exploration across different systems.
Moreover, it (iii) streamlines building data pipelines and defining access request processes by following the same standards across domains and sources. 
A federated model (iv) simplifies providing precise access control across sources and enforcing policies.
Furthermore, it allows to (v) assign non-overlapping roles to practitioners, and (vi) establish re-usable channels between data and analysts.
Finally, it could (vii) log the different analyses that other analytics teams have already performed such that others may use them (preventing work duplication).  
Nonetheless, while the federated model holds such advantages over the segregated model, we observed a similar analyst workflow (see Q7) and an adversarial position for the dataset request process.

\begin{comment}
\gmnote{Add small diagram explaining the central and distributed model including the roles.}
\end{comment}

% In the deep-dived organization, analysts could interact with data stewards and explore datasets through a user interface by performing limited SQL queries and using a preview function, which could only show a small number of rows.

\vspace{-1mm}
\subsection{Current Anonymization in the Industry}
\label{subsec:Anonymizing_Data}

In this section, we discuss the status quo of the anonymization that companies used to remain compliant without differential privacy, providing a baseline in the context of this work.

Companies can use data collected with user consent exclusively for the agreed \emph{primary} purpose.
If companies choose to use data for purposes other than the one agreed (\emph{secondary} purpose), data must be anonymized.
All companies had not deployed differential privacy in production or other advanced privacy-enhancing technologies, and employed traditional means of anonymization: \emph{suppressing} direct identifiers such as names, emails, or social security numbers, \emph{truncation} of, e.g., GPS locations and traces, \emph{generalization} (e.g., transforming $28$ into $[20, 30]$), and \emph{dropping} unnecessary attributes and outliers.
We consider these techniques \textit{syntactic}~\cite{syntactic_semantic_priv} because an algorithm transforms the data's syntax following a predetermined model (e.g., GPS locations must only have three decimals).
Additionally, data were always encrypted at rest. 

Beyond anonymization, to avoid merging multiple sources that could re-identify individuals, some companies did not allow analysts to access multiple datasets at once. 
In one company, depending on the purpose, stewards granted access solely to a subset of the dataset or a mock dataset for experimenting purposes.
Furthermore, for critically sensitive datasets (e.g., illnesses), one company provided access only to an anointed small set of analysts, limited access times, applied anonymization, and restricted analyses to cloud environments.
These environments produced logs for later auditing (if needed) and blocked analysts from downloading data.
On the other hand, based on user consent for primary use, analysts from one company could access detailed client profiles (names, house prices, mortgages, income, among others).
Despite having user consent, we recommend decoupling direct identifiers from the rest of the data (e.g., hashing the direct identifiers) to minimize the consequences of malicious analysts' actions, and encourage the integration of an automated process (or another practitioner) that can only access the analysis output and the direct identifiers to serve the customer (e.g., linked by a hash table only known to the additional process/practitioner).

Altogether, companies applied the principles of \emph{factual anonymity} (i.e., the effort of re-identification is disproportionate to the upside potential of an attacker learning about the individual), \emph{proportionality} (i.e., collection restricted to data necessary to fulfill the primary purpose)~\cite{architecture_of_privacy}, \emph{audit logging}, data sharing on a \emph{need-to-know basis}, \emph{data retention} and \emph{purging}~\cite{architecture_of_privacy}, \emph{access controls}, and traditional \emph{anonymization}.
However, the companies could not measure the privacy achieved by their systems and could only rely on their experience of what is compliant with regulation~\cite{GDPR_EU2016, ccpa2018}.

 \vspace{-2mm}
\subsection{Motivating Privacy}
\label{subsec:Motivating_Privacy}

\indent \textbf{RQ1:} \textit{What is the context of privacy protection in the targeted organization?}  

\vspace{0.5mm}
\noindent (\textbf{Q1}) \textit{What is the institution's motivation for privacy protection?} 
The five stewards agreed on two main motivations: (i) organizations have a legal and moral duty to abide by data-protection laws, (ii) privacy protection is an asset whose ``\textit{quality has to be equal to the premium product offered.}''

\noindent  (\textbf{Q2}) \textit{What are your privacy concerns when an analyst has full dataset access?} 
When proceeding with data protection risk assessments of dataset requests, stewards are predominantly concerned with misappropriation (i.e., unauthorized use of data) and data leakage.
While stewards do not expect analysts to be malicious, they are apprehensive of a potential lack of privacy skills, privacy-oriented mindset, and dataset understanding or pure negligence.
Specifically, stewards strive to prevent attacks such as unsolicited customer profiling, disclosing data to, or colluding with third parties to take advantage of the customer, combining datasets for re-identification, or using the data for purposes other than the one consented. 

\noindent  (\textbf{Q3}) \textit{At what level of data granularity are you protecting and measuring privacy?} 
The granularity of privacy protection is at the attribute level, and stewards measure privacy based on the fulfillment of data protection regulation.
For example, attackers could use the attribute \textit{location} to re-identify individuals; thus, according to GDPR~\cite{GDPR_EU2016}, the attribute must be obfuscated so that their home, work, and other points of interest cannot be linked to the individual.
Furthermore, the corporation must guarantee the ``\textit{security, transparency, and legitimacy of the [data] processing}.''
Overall, the anonymization approach strives to achieve the \textit{factual anonymity} principle. 

\noindent  (\textbf{Q4}) \textit{What could be improved in the dataset request process?} 
Data stewards suggested to (i) perform an audit to verify that the executed analysis aligns with the previous commitment, (ii) increase the quality of the datasets' metadata so that analysts can better define a purpose, (iii) increase the privacy training of analysts, (iv) produce privacy-enhanced dataset reports so that after the permission expires analysts can still retain some information, and (v) increment efforts in request process automation. 

\noindent  (\textbf{Q5}) \textit{What are your typical questions for the current interview-based full dataset access authorization? } 
To help other practitioners in the development of their risk assessment process, we gathered the most frequently asked questions from data stewards to data analysts during the dataset request process (see Appendix~\ref{app_sec:stewards_questions}).
Notably, without a clearly defined data usage purpose, the data stewards would not grant access to analysts. 

\noindent  (\textbf{Q6}) \textit{Instead of the interview process, would you be capable to run a program provided by the analyst such that the analysis is carried out without the analyst ever ``seeing'' the dataset?} 
% Lastly, we asked the stewards' opinion about executing a script provided by the analysts, who would obtain the outputs without ever accessing the dataset.
While most considered this an efficient, plausible, and necessary step in the future, the five data stewards did not yet have the required technical training, and their system did not enable the functionality.
``\textit{At the moment, it is not possible, but it will be a necessary step in the future, if not already today.}''
One steward remarked the importance of this functionality, as in some cases, e.g., requesting data from a branch of the company in another country, is extremely challenging.

 \subsection{The Practicality of Differential Privacy}
\label{subsec:The_Feasibility_of_Differential_Privacy}

\indent \textbf{RQ2:} \textit{Could differential privacy tackle the privacy-related pain points of an analysis workflow in an organization?}
\smallskip

\noindent  (\textbf{Q7}) \textit{What is your workflow to analyze data?} 
Despite the use of either a segregated or a federated model, the workflow was similar across organizations and employed common practices and tools; the main differences were in \textit{dataset exploration}.

\noindent (1) \textbf{Business use case demand}. 
    A business unit asked an analytics unit to conduct a study for supporting a business need, or analysts continuously studied data from a specific (customer's) domain. \newline
\noindent (2) \textbf{Dataset exploration}. 
    Only the companies using the federated model for accessing data could explore datasets' metadata through a \emph{data portal} without requesting access first (unless the dataset was tagged as critically sensitive), making the identification of the suitable dataset for the business need easier. 
    Analysts would find datasets using keywords in a search bar, and datasets provided descriptions, depicted their schema, and had data owners' contact information (analysts sometimes interviewed them to further understand the suitability of the dataset). 
     
    In the deep-dive organization, analysts could additionally perform any SQL aggregation query on the anonymized dataset prior to access (e.g., counts, averages, etc.), which they used for data understanding and quality checking (e.g., number of nulls and duplicates or measuring skewness).
    However, for privacy reasons, analysts could only retrieve a few rows when executing {\tt SELECT~*} query types and aggregations could time out (preventing excessive execution costs).
    Analysts used this \emph{preview} functionality frequently ``[...] \textit{to get a feeling for the data}'' and found it useful for exploration ``\textit{The preview query is the best feature.}''
    Companies without a federated model could not explore datasets, required data engineers for each use case, and analysts relied either on leveraging their contact network or on an experienced team lead to find promising datasets within the company. \newline
    % The analysts would then conduct interviews with the corresponding data owners to understand their data.
\noindent (3) \textbf{Dataset access request}. 
    Once the analysts identified a promising dataset, they formally requested access, which involved filling standard forms about the details and purpose of the analysis so that data stewards could assess the privacy risks.
    Except for three small companies, the request entailed interviewing with stewards, where they asked questions such as the ones in Appendix~\ref{app_sec:stewards_questions}. \newline
\noindent (4) \textbf{Visual inspection and preparation}. 
    With full dataset access, analysts would sometimes visually inspect the data values, types and schema.
    Analysts deemed these checks necessary because of the flaws sometimes found in the pipelines and dataset descriptions of the federated data portal or the data provided by the data owners in the segregated model.
    Moreover, as datasets consisted of many tables, analysts often checked which joins were possible and which attributes were most suitable for primary and foreign keys.
    With this information, they performed retrieval SQL queries with GROUP BY, WHERE, and {\tt JOIN} clauses to build a sub-dataset fine-tuned for their analysis.
    Many analysts also performed quality (double) checks and data wrangling using the Python's Pandas library~\cite{pandas} instead of SQL. \newline
\noindent (5) \textbf{Data analysis}. 
    Once analysts had checked the quality and wrangled the data, they primarily performed their analysis or ML model training in Python Jupyter Notebooks~\cite{PySpark}, and if the analyst dealt with \textit{big data}, they employed PySpark clusters~\cite{juypterNotebooks}.\newline
\noindent (6) \textbf{Output interpretation and model deployment}.
    If the use case required building a model for online prediction, the analysts would sometimes load the model into a more performant language like Scala before deployment.
    However, analysts frequently only needed to report statistics and visualizations, from which the business units drew actionable information. \smallskip
    
Most of the platforms and workflows employed AWS analytics tools~\cite{AWS} namely
S3 buckets (storage), Glue (data preparation), Athena (SQL querying), Sage Maker (data analyses), and analysts also used Python for visualization (one used R) and two of them complemented their results with Tableau~\cite{Tableau}.
Additionally, two analysts used Knime~\cite{Knime} for drag-and-drop analysis and visualization, and another two employed SAP data management software tailored to their department's needs.

The small interviewed companies had a few major differences, namely, they used a hybrid between the central (all datasets stored in a single data warehouse) and the segregated model.
Because of their small customer pool (managed centrally), they collected data from their customers or purchased user-data products from other companies to analyse or train ML models with more data, which required interaction with a segregated set of external data owners.
Furthermore, because of the small size of some companies, they had no need for formal dataset request processes as most employees were aware of the activities of the rest; their overhead was at the time of signing the initial contract with customers, which included data access policies and non-disclosure agreements.
They also employed traditional anonymization techniques and only retrieved with SQL data stored in, e.g., Google Cloud~\cite{GoogleCloud}, if strictly needed (less data for building the model and testing, and more data for the final training or analysis).
\smallskip

% We will discuss the implications of differential privacy in these steps in the next questions and in Section~\ref{sec:Discussion}.

\noindent  (\textbf{Q8}) \textit{Why do you need full dataset access?}
The main reasons given for accessing all the records of a dataset instead of, e.g., through solely a query interface, were:
\begin{enumerate}[label=(\roman*)]
\item \textit{Obtaining a holistic understanding of data}. All analysts worked uncomfortably if they could not make preliminary statistics or visualizations that encompassed all records ``\textit{I need to see the entire dataset to understand the data,}'' ``\textit{I am not necessarily sure of what I need to look at until I look at it. It is an improvisation, you start with a broad question and then you delve into it.}''
\item \textit{Less effort}.
A few analysts could fulfill their analysis with only SQL aggregation queries (e.g., counts and averages) and produced visualizations afterward; however, some found using other tools easier: ``\textit{Having access to the entire dataset allows me to use Pandas}.''
\item \textit{Cleaning data}. Given that there could be flaws in previous data preparation steps, analysts tended to (double) check all data for quality. 
\item \textit{Wrangling data}.
In the federated model, data engineers often built datasets without precisely knowing the purpose of a data analyst; thus, analysts sometimes took an engineering role, creating features for ML models or further tailor the dataset for their analysis by grouping or executing queries with {\tt JOIN} clauses. 
\item \textit{Debugging ML models}.
Analysts frequently needed to debug their ML models when testing and training, as there might be corrupted data points.
\item \textit{Visually inspecting values}. Some use cases, such as root-cause analysis, required analysts to check specific IDs and attribute values, and at times analysts needed to check whether an output table is feasible or map (truncated) GPS traces to street names for the analysis to be interpreted.
\end{enumerate}

Other analysts, however, did not always require access to all records because their ML model already converged, did not overfit, and provided enough accuracy: ``\textit{Since I am normally only doing exploratory work, I usually do not need access to the full dataset to prove that the given problem can be solved.}''
\smallskip

% \#2\#9
% they could use SQL, but it takes more effort with SQL than with python
% dislikes SQL beyond counts, no version control in AWS Athena, prefers pandas and for that he needs full dataset access
% \#3 \#4 \#5
% visual inspection of values to determine a root cause. check IDs
% \#6 \#13
% \#14
% understand the data
% \#7
% \#8
% data engineer For over-writing the dataset (new attribute, new data type) or performing a quality check to asses the quality of the data pipeline, e.g., looking for duplicates. 
% debugging when testing and training models because of data quality
% \#10
% location data needs to be mapped to streets
% \#11
% data engineering: for cleaning data you need to see all data
% data sc: explore whether data is skewed, make features for ML from time series
% \#12
% proving a problem can be solved with a ML model, no need access to the full dataset. only more data if the ML does not converge or not a good accuracy

\noindent  (\textbf{Q9}) \textit{How often do you request full dataset access? How long does it usually take?}
Among the large companies, the request frequency varied widely between $4$ times a month to once every $6$ months, with an average between once and twice a month.
Likewise, regarding waiting times, the minimum hovered around one to three days, while the maximum was two months, with an average between one to two weeks.
If another country hosted the data, the first request could take $9$ months.
Overall, analysts from the interviewed large organizations were blocked for at least one week for every new requested dataset, which they solicited on average once a month.
Specifically, in the deep-dive organization, analysts requested $5073$ datasets altogether in 2021 (around $14$ requests per day, which increased to $18$ as of 2022).
Out of all the requests in 2021, stewards rejected around $5.6$\%, amounting to fruitless weeks of revisions\footnote{The daily rejection rate went from $0.8$ in 2021 to $0.9$ in 2022, potentially indicating updated stricter policies.}.
Moreover, the number of requests was more than double the number of available datasets in the deep-dive organization in 2021 (a sign of significant duplication of work, accruing more costs).
On the other hand, three of the small organizations did not have such a formal request process, making them agile.
\smallskip

% \#1
% it takes 3 days, to weeks, to months, sometimes not allowed. normally 3 weeks
% \#2
% once a month
% \#3
% once every 3 months
% \#4
% once a week - usually 2 days, max 1.5 weeks if steward is on vacation
% \#5
% twice a month - 2 to 3 days, at most 2 weeks
% \#6
% once every 2 months - 2 days
% \#7
% continuous stream of user data
% \#8
% once every 6 months - 1 and 7 day, usually 1-2 days because trust has been built already
% \#9
% once every 2 months - it takes 2 months depending on the steward
% \#10
% once every 2 months, it takes between a week and an month
% \#11
% 2 per month - some days and a few weeks, it could take months if you do not know the process
% \#12
% once every 6 months - 4-6 weeks
% \#13
% once every 3 months - depends on the country, 9 months with contracts (money, GDPR), now it is one month
% \#14

\noindent  (\textbf{Q10}) \textit{What do you think about the process to request full dataset access in your organization?}
While analysts at small and US-based organizations were satisfied with the request process, there was an overall consensus at the EU-based large organizations on the following statement: ``\textit{The process to get customer data is slow. It might take from three days to weeks, to months}'' and for some, even ``\textit{Two to three days is too slow.}''
In the worst-case scenario, an analyst could wait weeks for a rejection.

Some analysts thought the interviews with stewards were primarily for building trust, and once built ``\textit{I always receive access. I do not see the point of waiting and interviewing every time.}''
Furthermore, frequently there were too many practitioners involved, leading to lengthy discussions about which dataset to use and often suffered a dilemma because responsibility entailed accountability in one organization ``\textit{If there is more than one data steward responsible, then it seems no one takes full responsibility for the acceptance or rejection of the request.}''
On the other hand, there were bottlenecks in the vacation season when only one steward was responsible.

Analysts agreed that accessing data has become better since they moved from a segregated model to a federated model; however, the process was still cumbersome, so much so that some teams incurred into the malpractice of entrusting a single analyst to manage the process.
One analyst summarized the inefficiency of the segregated model:
``\textit{There is a lot of bureaucracy and everyone is extremely reluctant to grant access to a full dataset. Even for internal problems and non-sensitive data. It is cumbersome to request full dataset access because there is no central point where the dataset access can be requested and no central entity which manages access control and usage control for all datasets. For every instance, the process is a bit different depending on the responsible department, underlying workflow and data pipeline.}''
In the segregated setting, the process was lengthier, and an analyst could not explore what others had analyzed or requested, sometimes leading to redundant work.
\smallskip

\noindent  (\textbf{Q11}) \textit{What features do you think are missing in your organization's data analysis workflow?}
The most notable proposed improvements were: (i) including rich information regarding dataset metadata (preferably with visualizations) and their access request process, (ii) improve real-time analytics performance, (iii) enabling full analytics in data portals such that an analyst does not need to transfer data to other tools, (iv) limiting access times to improve security, and, from a data engineering perspective, (v) automating sensitive data detection and (vi) improving quality and automated checks to minimize visual inspections.

\begin{comment}
a data engineer mentioned that ``\textit{there should be a distinction between read and write access}.'',
a problem of interoperability between two data portals in the same organization.
enable arbitrary library installation in the provided tools, 
upload datasets
\end{comment}

\begin{comment}
\#1
data sc. in the portal (apply the same analytics to another dataset), it is not convenient to jump to AWS
more metadata information
\#2
pre-defined libraries for data science in the available tools, cannot install whatever you want
one analyst with two use cases, both use cases need the same dataset, he needs to make 2 requests.
testig data is different to production data
\#3
limit access times, it only depends on the analysts to cancel the access
\#4
more information about datasets
if one creates a dataset from another dataset, the portal does not allow you to upload it
\#5
Improve UI
data access request should be guided
\#6
allow to explore data through visualizations
\#7
more checks on data quality, with high quality, then no need to look into the data
"There is a still a ways to go to deploy DP."
\#8
information on how data is joined
quality measurements sensitive data detection automation
build semantic layers and visualize them
\#9
--
\#10
better metadata about datasets
\#11
more quality data about datasets
\#12
too many workflows
\#13
not agile exploring new tech and their visualization
\#14
data visualization
\end{comment}

\smallskip
\indent \textbf{RQ3:} \textit{When does differential privacy impede an analysis? }
\smallskip

\noindent  (\textbf{Q12}) \textit{In which analytics use cases have you been involved?}
Most analysts worked on \emph{descriptive} use cases.
Some of these use cases focused on reporting conclusions from the past by performing root cause (error), cost down, and warranty costs analysis.
Other analysts strove to increase the situational awareness of the company by analyzing location-based time series of users (identify points-of-interest or common traces), their behavior when using a product or a service (frequently used features, A/B tests, purchases or component performance), and demographics (user-base analysis or advertisement).
Additionally, some analysts focused on alerting internal stakeholders of quality defects in real-time, and another analyst performed correlation analysis to better understand the interplay of different variables in products and services.
Most of these use cases required performing aggregate statistics (including visualizations to report to management), namely for situational awareness, while others demanded visually inspecting exact values (namely for error detection or financial data), and one analyst used classification ML models for quality checks.

The minority of interviewed analysts were involved in \emph{predictive} use cases: forecasting product lifetime, labelling spam and inappropriate images, user behavior, the company's profit and loss, claim costs, and predictive maintenance and creating automated underwriting models.
While these use cases relied on basic statistics, some used vanilla ML such as linear regression (for underwriting models).
Nonetheless, the interviewed analysts agreed that using ML was rare; thus, most analysts relied on aggregation and visualizations, as the business units demanded \textit{quick} and easily \textit{interpretable} results.
\smallskip

\noindent  (\textbf{Q13}) \textit{Is SQL-meaningful for your work? How many SQL-like queries do you make weekly?}
Most of the interviewees employed SQL, chiefly during exploration, and they deemed SQL an important part of their workflow ``\textit{SQL is amazing, everyone who tells you SQL is going away is wrong},'' because they could quickly look into rows and performed preliminary statistics, and, with {\tt JOIN} clauses, prepare a dataset for their use case.
The least adept analyst executed $5$ weekly queries, while the most assiduous SQL user performed $250$, being the average around $50$ queries per week.
\smallskip

\begin{comment}
50 daily queries in average.
Depends on the number of datasets he deals with. It ranges between 5 and 50 weekly.
5 per day.
For exploring only and right before pre-processing. 1 query a day.
1 every day. 
Currently, none
2 SQL queries per day.
25 per day.
4 a week
100 per week
4 a day
\end{comment}

\noindent  (\textbf{Q14}) \textit{How often do you need machine learning to fulfill your analysis in contrast to using SQL?}
Two interviewees always needed ML to fulfill their analysis, while another $4$ used ML for some of their use cases.
The analysts who were allowed to explore datasets used SQL for exploration, and three used SQL to generate statistics and completely fulfill their analysis (complemented with visualizations), while the rest preferred Python or other tools for analysis.
Furthermore, analysts often visualized data to accompany their results with other tools (see Q7) and employed retrieval SQL queries for visual data inspections (e.g., for error analysis) or building tailored datasets for their analysis.
%using Python, Tableau~\cite{Tableau}, or Knime~\cite{Knime}, and employed either Python's pandas library~\cite{pandas} or retrieval SQL queries for visual data inspections (e.g., for error analysis) or building tailored datasets for their analysis.
\smallskip

\noindent  (\textbf{Q15}) \textit{What are your most used machine learning models?}
The $6$ analysts employing ML most often resourced to decision trees and linear regression because they are easy to debug, interpret and visualize the results.
These analysts also mentioned the use of random forests or XGboost (preferred), Bayesian approaches, support-vector machines, and, for time series, they used outlier detection techniques for error analysis and autoregressive integrated moving average for forecasting.
Analysts avoided neural networks because they are hard to interpret; nonetheless, one practitioner indicated they were working on deploying neural networks in the future for underwriting models.
In particular, one analyst employed PyCaret~\cite{PyCaret} for automated ML workflows, as in the corresponding department ``\textit{It is more important to be quick and give a good-enough overview than having well trained precise models,}'' ``\textit{Complex machine learning is often never required}.''
Other analysts voiced that such is often the case.
\smallskip

\noindent  (\textbf{Q16}) \textit{If you were to use differential privacy to fulfill your analysis, when and how much accuracy would you be willing to forgo?} 
The willingness to forgo accuracy depended on the use case, with a spectrum ranging from the need for absolute accuracy for quality, error, or financial analyses, to indifference for accuracy in exploratory use cases (only enough accuracy to prove a solution works).
For the rest of the use cases, while the interviewees would need to estimate the minimum accuracy formally, they informally reported on average that an accuracy of around $98$\% would be sufficient, and none reported below $95$\%.
Some financial analyses could also allow errors in the magnitude of cents of a monetary unit, and one analyst reported the need for at least $99$\% accuracy for finding suitable primary keys for joins. 
Additionally, comments such as ``\textit{I am scared of introducing noise into the analysis. [...] \textit{From all the analyses I do every year, there will be some that will be wrong.} [...] \textit{How well you are compensated depends on how well you do.} [...] \textit{Because you are paid to have an opinion, you are not allowed to be wrong},'' suggest that organizations' incentive systems for data scientists, e.g., bonuses, should change to account for errors due to differential privacy.}
\smallskip

\indent \textbf{RQ4:} \textit{How would differential privacy affect the workflow of an analyst?} 
\smallskip

\noindent  (\textbf{Q17}) \textit{How much would the noise affect your analysis?}
Depending on how much the noise could affect an analysis, we observed three categories for use cases: (i) suffer adverse effects, (ii) reach a tradeoff, and (iii) robust to noise.
The first one relates to analyses reporting error, quality, or financial results, where noise could have catastrophic consequences, e.g., a defective component is installed in a product, or yearly budgets are inflated.
Moreover, analysts sometimes dealt with low data quality (notably from sensors) that noise could worsen, e.g., GPS locations may already have a $10$m error, making a points-of-interest analysis noisy in itself. Adding noise to the aggregation might produce unusable results.

The second type concerns aggregation and visualization reports, where, given enough data, the noise would not affect the interviewees' analysis workflow (e.g., demographics or product usage studies); however, analysts would prefer working with error bounds to report confidently to management.
The third type of noise relates to analysts testing solutions ``\textit{Since my work is exploratory and we mostly try to prove that the problem can potentially be solved, noise would not have any negative effects for my analysis.}''
\smallskip

\noindent  (\textbf{Q18}) \textit{Would you find it helpful to execute differentially private SQL queries to explore and fully analyse datasets without the standard permissions?} 
We theorized that given the plausible deniability guarantees of a differentially private analysis, which can be argued to comply with the identifiability notion in GDPR~\cite{GDPR_EU2016, GDPR_DP}, some uses cases that heavily rely on aggregation might abate or not need the standard dataset request processes.
From this perspective, most analysts found differentially private SQL queries helpful, in summary, because ``\textit{If having differentially private SQL queries for data exploration implies reduced bureaucracy and easier access, then this would save a lot of time and discussions.}''

Notably, one interviewee saw the potential of differentially private queries for data exploration: companies could expose data externally through an API, allowing others to understand their data products by conducting preliminary analyses.
Another analyst favored integrating differential privacy into, e.g., AWS Athena~\cite{AWS}.
On the other hand, few analysts did not see the value of differential privacy because their use cases required, e.g., visual inspections for error detection, or their organizations were already agile in accessing data. 
Lastly, two analysts voiced a general concern ``\textit{[Differential privacy] is a double edge sword. You could get quick [data] access, but then [results are] noisy,}'' ``\textit{I think I would find it annoying, since it adds an additional step and obfuscates the results},'' and it could lead to confusion as analysts usually work with accurate data. \newline 
\vspace{-2mm}

\noindent  (\textbf{Q19}) \textit{Only based on the information extracted from a dataset exploration~with differential privacy, could you write a script to fulfill your analysis goal?}
% With this question, we aimed to assess the confidence of the interviews to write scripts without full dataset access by relying only on differentially private query outputs, thereby avoiding lengthy request access processes.
A couple of interviewees shared their inability to program their script as they needed to see the data (e.g., error analysis), and the others shared their skepticism by highlighting the problem of low data quality.
Even if an analyst developed an intuition for the data through differentially private aggregation queries, programming other statistics, visualizations, or ML models would likely require debugging, which may lead to visual inspections.
\smallskip

\begin{comment}
No. He needs to see the data and clean it beforehand.
No, because data quality is most of the time bad, and, therefore, data needs cleaning, which can usually only be done by looking at the data. With only having an intuition of what the data contains by means of DP, the script they would produce would fail most likely because of ingesting un-cleanded data.
In his case he cannot. 
The scripts he builds are SQL queries to join datasets in the data pipeline, so he would need to perform exploratory retrieval SQL queries to know whether a join is possible. Even if SQL is enhanced with a DP mechanism for element selection, e.g., with the exponential mechanism, the answer is noisy, so the data engineer might think a dataset can be joined when it cannot, or viceversa.
Yes, it would be possible - but for debugging it might be hard.
No, because of data quality - needed for debugging.
He needs the preview, because then he can see the content. Otherwise he cannot. 
Depending on the difficulty of the task. It is definitely possible to get an idea, if the dataset is the right choice and see, if the direction we’re going is the right one. By this, we could see if it is even worth going through the process of getting the full dataset.
No, difficutl without minimal information. 
It would not work because he needs to find out the primary keys. For the 10% of the imte where he is doing counts and averages 
count distinct values / count of all values <- if he gets 99%, than that is the primary key
\end{comment}

\noindent  (\textbf{Q20}) \textit{What are the minimum properties for you as an analyst such that you are confident to write an analysis script without full dataset access?}
Assuming enough data quality and a use case that does not require visually inspecting data, the interviewees indicated that for tentatively writing code without dataset access, they needed: good metadata from the dataset, such as attribute descriptions, knowledge about the events that trigger data collection, primary keys, data types (IDs, dates, timestamps, floats, strings), dataset size (number of rows and columns), and attribute distributions to learn about sparsity in the form of histograms or box plots.
\smallskip

\begin{comment}
Perhaps if there is a lot of information about the data, many pages of good documentation per dataset.
As long as the data is clean and one can execute the aforementioned queries such as rolling averages, counts, and std, then writing the script would be possible.
In order to write a script to join datasets, he would need to perform a metadata analysis: Averages, min and max of all columns, percentage of nulls a column contains, access to the schema to know whether the join will work, data types.
Number of updates, number of attributes collected, is there any pre-processing of the values, data types, the partitions. 
She would need to know that the data has high accuracy, data schema. 
The dataset schema, size of the dataset (number of columns and rows), type of data in each column (IDs, dates), distribution information (histogram of each atribute, so they know aout sparsity - he likes the Kaggle data profiling)
Access to the data structure, type and amount of data points. Mock data or a sample would help.
if you can get Max, min, median, what python describe tells you. Box plots for distirbutions per column. quaertiles.
He does not see how his job would be done
\end{comment}

\noindent  (\textbf{Q21}) \textit{Would you find it helpful to use a dynamic dashboard that visualizes dataset information with differential privacy?}
Since data platforms may not expose sensitive data on a dashboard for exploration, we conceptualized enabling this functionality with differential privacy.
All but one interviewee considered such a dashboard helpful for finding a suitable dataset faster and with a better user experience than their available utilities (static and scant summaries or using SQL).
Specifically, an interviewee commented that, in general, one should be able to visualize the data and get basic statistics before requesting access, and another analyst would have liked to preview similar information as the ``{\tt describe}'' method of a Pandas dataframe~\cite{pandas} (count, mean, standard deviation, minimum, quartiles, maximum).
Nonetheless, one analyst noted that a dashboard is a nice-to-have because it is only more convenient than SQL.
Lastly, another interviewee underlined a problem that may arise when an analyst does not trust the data provided by the visualization, e.g., when the plot seems implausible.  
The interviewee suggested that a dashboard should enable the analyst to drill down or provide contact information from a data owner to verify correctness.
\smallskip

\smallskip
\indent \textbf{RQ5:} \textit{Can differential privacy enhance the privacy of the frequent SQL-like queries analysts execute? }
\smallskip

\noindent  (\textbf{Q22})  \textit{What are your top SQL-like queries before you have full dataset access?}
If analysts could explore datasets, most would usually conduct a metadata analysis with SQL to assess data quality: finding the number of duplicates, outliers, nulls, and not-a-number values and measuring the skewness.
Analysts would also explore the dataset for data understanding using {\tt COUNT}, {\tt DISTINCT},  {\tt MAX}, {\tt MIN}, {\tt AVG}, and {\tt VARIANCE} functions with {\tt WHERE} and {\tt GROUP BY} clauses.
Analysts were typically interested in frequent values within a column (see details in Appendix~\ref{app_sec:frequent_queries}).
Furthermore, the deep-dive organization allowed to use {\tt SELECT~* LIMIT(X)} for a few {\tt X} rows so that analysts could have a ``\textit{feeling}'' for the data.
%Moreover, in their data portal, three of the top five upvoted SQL functions were {\tt COUNT}.
On the other hand, fewer analysts performed retrieval queries (limited in output rows) to verify whether an ID was present or two tables could be joined.
\smallskip

\noindent  (\textbf{Q23}) \textit{What are your top SQL-like queries after you have full dataset access?}
Analysts who could not explore the dataset prior to having dataset access would execute queries such as those in Q22 first (see Appendix~\ref{app_sec:frequent_queries} for details).
Afterward, if they did not already retrieve the necessary information from the exploration, they resorted to Python and other visualization tools to fulfill the use case.
Some analysts performed additional retrieval SQL queries with {\tt JOIN} and {\tt SELECT~*} clauses with different filters to visually inspect data points (e.g., IDs or potential errors), identify cut-offs (e.g., where an attribute data type changes), or fetch the specific data they needed.
\smallskip

\noindent  (\textbf{Q24}) \textit{What is the ratio between aggregation queries and queries to retrieve items?}
While the interviewees would need to calculate the percentage formally, they reported informally, on average, that around $30$\% of their queries were for aggregation, being the lowest $0$\% and the highest $90$\%.
Another three analysts used SQL for retrieval and Python for aggregation or vice versa.

\begin{comment}
aggregation
80
10
30
25
20
25
90
0
10
60
20
1

99

20 - pyhton for retrieval
0 - SQL for retrieving
80 in python

(v2)
80% aggregation vs. 20% retrieve item.
"20% aggregation vs. 80% retrieval (if we consider retrieval analysing data with Jupyter Notebooks). He only uses SQL queries to get aggregated values. To receive items, they use Glue Jobs or Sage Maker.
"
90% retrieve items vs. 10% aggregate queries.
1 aggregation vs. 2 retrieval
With Python. 80% aggregation vs. 20% retrieval.
SQL: 100% retrieve items. Considering Python and SQL queries, 50 % aggregation with Python vs. 50% retrieval with SQL.
25% aggregation vs. 75% retrieval
80 % retrieval vs. 20% aggregation.
75% retrieval vs. 25% aggregation
"Data engineering: 90% retrieval vs. 10 aggregation
Data scientists: 90% aggregation vs. 10 retrieval"
Don’t know
SQL: Retrieve items 100%. 
1 in 10 are aggregation
\end{comment}

\section{Discussion}
\label{sec:Key_Findings}

In this section, we present selected key findings (KF) distilled from the data stewards' and analysts' answers to the $24$ interview questions of section~\ref{sec:Results_of_the_Case_Studies}, most accompanied by succinct recommendations. 
Lastly, we answer the research questions proposed in section~\ref{sec:Research_Method}.

\subsection{Key findings}

\noindent \textbf{(KF1)} \textit{Data stewards seem to be more concerned about security than privacy.\footnote{In the context of this work, we refer to \emph{security} as the measures for blocking \emph{unauthorized} data access, while \emph{privacy} focuses on limiting harm by \emph{authorized} analysts~\cite{architecture_of_privacy}.}}
Data misappropriation and leakage retain the most attention (Q2), which is reflected in the established cumbersome dataset request processes that dictate access controls and accountability in the name of building trust with analysts.
However, we highlight that privacy lacks such attention, even when some companies still allow their analysts to ``\textit{see}'' or download sensitive data.
Attacks on privacy (Q2) could enable malicious analysts to misuse data for spying on or leaking secret information of celebrities, acquaintances, ``friends'', or relatives~\cite{celeb_priv}, blackmailing, or discriminating individuals in social or commercial transactions online~\cite{privacy_economic_good}.
Despite the risks, companies predominantly use traditional and potentially vulnerable anonymization techniques (e.g., pseudo-anonymization or k-anonymity), as demonstrated by the research community~\cite{dwork_exposed_2017, sweeney_identifying_2013, gao_elastic_2014, kondor_towards_2020, narayanan_robust_2008, archie_anonymization_nodate}.
Thus, we suggest companies increase efforts to research and deploy more advanced PETs. \newline

\noindent \textbf{(KF2)} \textit{Running analysts' scripts without ``seeing'' the data is a distant reality for the interviewed companies.}
We explored multiple ways for analysts to run scripts without direct dataset access.
In Q6, stewards declared their technical inability to execute scripts that analysts could share and, thus, avoid granting them access (saving time).
With the proper tooling, a non-technical steward could potentially run the script; however, current systems do not offer such abstracted functionality and this option would relay the responsibility to the stewards instead.
An alternative to transferring the trust to stewards consists on executing analytics in trusted execution environments\footnote{Hardware and software designed to run applications securely against unsolicited retrieval of sensitive information or key material~\cite{sec_HW_1}.}.
Additionally, analysts altogether gave six reasons why they needed full dataset access (Q8) and reported skepticism when asked about writing a script (beyond aggregation) based on a differentially private exploration (Q19 and Q20).

The main impediment reported was data quality, which often led to visual inspections of dataset values.
As encouraged by point Q11-(v), we suggest companies prioritize increasing data quality as it will indirectly improve privacy and increase the technical training of data stewards and owners, enabling more data security options. \newline

\noindent \textbf{(KF3)} \textit{Given the analysis workflow, differential privacy could have a significant impact on dataset exploration} (see Q7).
As long as exploration does not require visually inspecting a particular ID or an exact attribute value, differential privacy can provide noisy statistics for the analyst to familiarize with the data (e.g., number of rows, averages, quantiles, etc.), which is often enough to assess the dataset's suitability.
Furthermore, while analysts were not allowed to explore critically sensitive datasets with SQL, employing differential privacy could arguably enable their exploration by adding an extra layer of protection.
Additionally, platforms could provide privacy-enhanced dataset previews (e.g., only revealing a few rows or producing dummy or synthetic data with or without differential privacy).
Overall, differential privacy could facilitate exploration that otherwise might not be possible or timely. \smallskip
%, and, with enough accuracy, an analyst can find suitable primary keys by checking the (noisy) percentage of distinct values

\noindent \textbf{(KF4)} \textit{Analysts could employ differentially private mechanisms to fulfill certain use cases} (see Q7).
If the analysis requires summary statistics and visualizations, a differentially private analysis could fulfill the privacy-utility tradeoff given enough data. Consequently, analysts could fulfill use cases without exact outputs, avoiding potential privacy leaks.
Regarding ML, while its differentially private implementations are at an early stage, researchers and practitioners could explore systems to assess whether a model shows signs of converging with enough accuracy after training on a sample of the target dataset.
Such a system could help analysts to determine the validity of the model or the dataset.
Lastly, we suggest exploring whether differential privacy can enable more accurate analyses than the current organizations' anonymization processes. \smallskip
%In summary, introducing differentially private could streamline the workflow of analysts by providing exploration for formerly inaccessible (critically sensitive) datasets and, in some cases, fulfilling analyses without direct data access.

\noindent \textbf{(KF5)} \textit{After fulfilling the use cases, the interviewed companies do not have a human-supported privacy auditing step.} 
The last reported step of the workflow in Q7 was ``\textit{output interpretation and model deployment}''. Aligned with a steward in Q4: ``\textit{perform an audit to verify} [\textit{alignment with analysis commitment}],'' we suggest privacy officers in companies add a randomly-sampled auditing step with a human in the loop after the conclusion of the use case.
We also suggest audit logs, which one of the interviewed companies produced for every execution on sensitive data in secured machines, where analysts could not download data or install new software. \smallskip

\noindent \textbf{(KF6)} \textit{Given the six reasons analysts shared for fully accessing datasets, differentially private mechanisms could help in (i) ``obtaining a holistic understanding of data'' by providing dataset summary statistics} (see Q8).
Additionally, we suggest substituting tedious SQL analyses with dashboards for visualization, so that tasks require (ii) ``\textit{less effort}''.
We also suggest engineers develop and integrate tools that enable analysts to (iii) ``\textit{clean}'' and (iv) ``\textit{wrangle data}'' without visually inspecting the values (i.e., no complete data access required). 
With such tools, filtering values, imputing, removing duplicates and outliers, fixing wrong formattings, handling missing data, or creating new attributes would also help with (v) ``\textit{debugging ML}.''
Moreover, aligning data engineers with analysts could improve data quality, e.g., by involving engineers in the conversations between stewards and analysts.
Lastly, researchers could investigate how differentially private set union mechanisms~\cite{dp_set_union_1, dp_set_union_2} could help analysts to (vi) ``\textit{visually inspect values}.'' Meanwhile, we suggest increased security measures for such cases.
\newline

\vspace{-1mm}
\noindent \textbf{(KF7)} \textit{Analysts are frequently blocked for significant periods every time they request access to datasets} (see Q9).
There are a few consequences of such delays.
Data stewards and privacy officers must also invest their time in reviewing the requests. 
From our conversations with the interviewees, we also learned that long waiting times could hamper analysts' bursts of creativity and productivity, which indirectly negatively affect the quality of work.
Additionally, an interviewee recounted the malpractice of deferring all the dataset request process responsibility to a single analyst in the team (see Q10). 
Such practice overburdens an individual with the responsibilities of the entire team for, e.g., a data leakage, creating an unhealthy imbalance in accountability. This practice further increases the company's privacy risk by potentially having the other analysts handle data without privacy training.
We suspect this malpractice is a sign of over-complicated dataset request processes and long waiting times; thus, we suggest privacy officers streamline their processes and prompt teams to refrain from overburdening a single analyst.

Differential privacy's stronger guarantee could reduce the complexity of the interactions between practitioners by offloading their data protection demands and, thus, reduce the costs accrued by these human-intensive processes. 
Lastly, given that there were multiple requests for the same datasets from different teams, we encourage companies to build interfaces depicting privacy-enhanced summaries of past fulfilled use cases per dataset. An example is the repository designed by Johnson~et~al.~\cite{share_analyses} in the health industry. \newline

%Taking an average of one-week waiting time for each of the $5073$ requests, it accrues to an estimate of eight years of stewards and privacy officers assessing requests and analysts waiting for access every month\footnote{Note that analysts, and stewards and privacy officers perform other tasks in parallel.}.

%Nonetheless, \emph{optimistically} assuming they all spent $10$\% of their time on filing or revising requests, respectively, their combined monthly work would accrue to approximately ten months, which, for one business year, it is equivalent to ten yearly salaries}% of ten practitioners focused on filing and revising requests.}.
\vspace{-1mm}
\noindent \textbf{(KF8)} \textit{Differential privacy could arguably reduce the time to access data.} 
As differential privacy brings a higher and formal guarantee of privacy, it could relax the inquisitiveness of data stewards, eliminate (steps of) the request process, and enable exploration that was otherwise not possible.
By enabling exploration, analysts reduce the likelihood of investing time in request processes that could even result in accessing a non-suitable dataset.
With exploration and higher privacy guarantees, differential privacy could also speed up requesting data from other countries, which seemed the most significant bottleneck (see Q9).
Additionally, differential privacy could potentially prolong access times (if these are limited) and shorten development cycles with an earlier data access by testing algorithms and applications with noisy data or outputs.
Regarding applications specifically, once finished, the customers can confirm whether the product works appropriately with real data.

We have also observed that, once analysts have dataset access, much of the data protection and accountability lies on their shoulders, which differential privacy could lift to a degree by protecting beyond trust and policy.
However, the analyst somewhat familiar with differential privacy pointed out that, unless data quality is improved (as also suggested in Q11), "\textit{There is a still a ways to go to deploy differential privacy,}" because the need to debug by visually inspecting data will prevail. 
To increase privacy protection in those cases, we suggest using differential privacy with limited visual inspection. \smallskip 
%Because differential privacy can provide metadata information (e.g., number of records, nulls, duplicates, skewness values, among others) for critically sensitive datasets (formerly restricted for exploration) in a privacy-compliant manner, which

\noindent \textbf{(KF9)} \textit{Most analysts employed aggregations and visualizations to fulfill their use case in a timely manner, while machine learning was not as predominant} (see Q12).
We found that analysts could employ differential privacy to explore datasets suitable for all the identified use cases.
However, for the analysis itself, the interviewees voiced that the noise would invalidate the use cases related to quality, error, and (some) financial analyses because mistakes in safety decisions and financial planning are company critical.
Nonetheless, for the use cases that required aggregation and visualization, with enough data, we suggest analysts fulfill these use cases with differentially private queries such as counts, averages, and percentiles, among others (e.g., user behavior, demographics, and some location-based analyses).
However, the available tools for differentially private ML are not mature for widespread adoption. Thus, we encourage researchers and practitioners to improve and build systems around existing proposals in future work, e.g., location-date analysis~\cite{localization_DP}, heavy hitter identification~\cite{heavy_hitter_dp_permute}, mining frequent itemsets~\cite{dp_itemsets_mining}, deep and supervised learning, random forests, and linear regression, among others~\cite{DP_DL, privgene, dp_random_forest, IBM_repo, causal_graphs_dp}. \newline

\vspace{-2mm}
\noindent \textbf{(KF10)} \textit{For the interviewed companies, SQL was more important than machine learning and was considered a meaningful tool frequently employed in their workflow} (see Q13 and Q14).
Additionally, on average, $30$\% of the top SQL queries executed before and after full dataset access were for aggregation (see Q22, Q23, and Q24), which researchers have already adapted to fulfill differential privacy~\cite{Uber_DP, Chorus_repo, ZetaSQL}.
Therefore, there is still a gap between what researchers have enabled and what practitioners need for enhancing the privacy of their frequently used SQL queries---a gap we intend to partly cover in section~\ref{sec:Query_Re_Writer} by proposing $10$ key system desiderata that an integrable privacy-enhancing analysis system should fulfill.
Beyond SQL, differential privacy and its available tools are also suitable even when analysts preferred using Python for aggregation and ML use cases that allowed for lower precision.
In particular, we encourage using Python libraries such as IBM's diffprivlib~\cite{IBM_repo, DP_benchmark} that provide many off-the-shelf differentially private ML models that could provide enough precision for the intended purpose, such as for the linear regression model one company used for underwriting (see Q15).
However, practitioners will require further engineering to limit Python to strictly privacy-enhancing libraries and amenable standard functionalities (e.g. by using policy enforcement paradigms such as Wang~et~al.'s Data Capsule~\cite{data_capsule}). \newline 
%(e.g., linear and logistic regression, random forests, PCA, among others).

% REDUNDANT: However, after exploration, some analysts shifted to Python to perform the analysis even though they could retrieve the statistics needed with SQL for some use cases; these analysts considered SQL not as intuitive as other tools.
\vspace{-2mm}
\noindent \textbf{(KF11)} \textit{Analysts confirm that differential privacy would be helpful for dataset exploration, fulfill certain use cases, and for enabling privacy-enhancing dashboards for dataset visualization} (see Q18 and Q21).
For aggregation-based use cases where noise has no detrimental effects, analysts informally reported, on average, a required accuracy of $98\%$ (see Q16 and Q17).
While such a figure might seem high, given the large amount of data handled, analysts could potentially find enough for aggregations that fulfill their privacy/utility tradeoff.
For example, as of early 2022, the deep-dive organization had roughly $2260$ datasets in its federated system amounting to $3.4$PB ($1.5$TB per dataset on average) with an average daily query execution of over $900$TB.
However, size might not be enough for some use cases, as the analysis could be sensitive to outliers or corrupted data.
Lastly, we observe that it is critical for analysts to know whether the accuracy is above their required level, which would consume privacy budget and be hard to estimate, e.g., when the analysis needs post-processing (clamping or truncation). \newline

\begin{comment}

Moreover, a predominant problem today is poor \emph{data quality}, which often forces analysts to ``\textit{see}'' the data for debugging code, thus preventing the emergence of new systems that enable analyses without direct data access.
Overall, We regard enabling analysts to work without ``\textit{seeing}'' data as a critical, multifaceted challenge for the research community to solve, and we consider differential privacy a building block for the endeavor.
\end{comment}

%\vspace{-1mm}
\subsection{Answers to the Research Questions}

\textbf{RQ1:} \textit{What is the context of privacy protection in the targeted organization?} 
The deep-dive organization invests more resources to security than privacy-enhancing analysis---pattern also present in the other organizations.
Moreover, stewards consider privacy an asset and strive to provide the best standard for their customers. 
However, organizations still employ traditional anonymization techniques.
Furthermore, companies today are unable to tangibly measure the privacy of their process (see Q3), and while there are specific privacy and security measures such as access controls, they are hard to quantify formally.
%On the other hand, differential privacy can provide such measurement for some steps of the analysis workflow.
%Nonetheless, there is a lack of consensus in the community about what should be the correct value of $\varepsilon$ in different contexts~\cite{dwork_differential_2019}.
Lastly, we observed that the interviewed companies are far from having ``\textit{data at their fingertips}'', one of the reasons being the onerous dataset request processes, which confuse access hardship with access protection.
\smallskip

\noindent \textbf{RQ2:} \textit{Could differential privacy tackle the privacy-related pain points of an analysis workflow in an organization?}
Yes, to a large extent---In essence, the main problems are (i) lengthy and cumbersome dataset request processes.
Moreover, given that analysts can sometimes ``\textit{see}'', download, and share the data once they are granted access, and even collude with other co-workers with access to linkable datasets, (ii) \emph{only} policy protects data once stewards grant access.
Based on our work, we argue that differential privacy can reduce time-to-data by enabling exploration of critically sensitive data or across third-party data sources, relax the current data access restrictions thanks to its formal privacy guarantee, is applicable to some aggregation-based use cases, and, for some use cases, engineers should consider building solutions that block analysts from ``\textit{seeing}'' the data. 
\smallskip

\noindent \textbf{RQ3:} \textit{When does differential privacy impede an analysis? }
The answer to this RQ heavily depends on the use case and whether the analysts are willing to forgo accuracy. 
On the one hand, noise addition-based differential privacy is useful in aggregations performed by the interviewees (e.g., querying demographics or frequently used product features). Moreover, on average, interviewees were comfortable with $98$\% accuracy.
However, differential privacy is not a silver bullet, as some of the interviewees' use cases cannot rely on it (e.g., error analyses or critical financial estimations).
Therefore, we suggest building systems that enable differential privacy while maintaining the flexibility of allowing non-differentially private queries when the use case strictly needs them.
%On the other, the noise added to, e.g., battery error analyses or critical financial estimations, could have catastrophic consequences in, e.g., safety or financial planning.
%Overall, there is a large number of aggregation-based use cases that can benefit from differential privacy.
\smallskip

\noindent \textbf{RQ4:} \textit{How would differential privacy affect the workflow of an analyst?} 
If differential privacy enabled previously unavailable exploration and provided data for privacy-enhanced dashboards, analysts would have a better user experience in their workflow and lower time spent on processes and exploration, but would also need to accustom to working with noisy data.
\smallskip

\noindent \textbf{RQ5:} \textit{Can differential privacy be applied to the frequent SQL-like queries analysts execute?} 
Yes---While not as frequent as retrievals, around a third were aggregations amenable to differential privacy. \break

\section{Towards Practical Differential Privacy}
\label{sec:Query_Re_Writer}

This section provides a set of critical system desiderata a differential privacy (DP) analytics system should satisfy for practical deployments.
Subsequently, we identify requirements fulfilled by state-of-the-art tools (see Table~\ref{tab:mapping_tools_requirements}) and highlight the gaps in practice.

\subsection{Key System Desiderata}
\label{subsec:desiderata}

%%%%%%%%%%%%%%%%%%%%%%%%%%%%%%%%%%%%%%%%%%%%%%%%%
%%%%%%%%% REFACTORED TABLE %%%%%%%%%%%%%%%%%%%%%% 
%%%%%%%%%%%%%%%%%%%%%%%%%%%%%%%%%%%%%%%%%%%%%%%%%
% - Original table commented at the end

\begin{table*} [htpb!]
\centering
\small

\vspace{-5mm}
\begin{tabular}{ |M{2.3cm}|M{0.95cm}|M{0.95cm}|M{1.2cm}|M{1.1cm}|M{0.75cm}|M{1.2cm}|M{1.4cm}|M{1.4cm}|M{1.3cm}|M{1cm}| } 

\multicolumn{11}{c}{\textit{Table 1A: Libraries, frameworks, and systems for differential privacy analytics.}} \\

\hline

 & \textbf{(I)} & \textbf{(II)} & \textbf{(III)} & \textbf{(IV)}  & \textbf{(V)} & \textbf{(VI)} & \textbf{(VII)} & \textbf{(VIII)} & \textbf{(IX)} & \textbf{(X)} \\ \cline{2-11} 
\centering  \textbf{Tool/Desiderata}  & DP \break Analytics & Usability & Security & Synthetic Data & Visuals & Privacy Budget & Accuracy Adjustment & Query Sensitivity & Data \break Annotation & Access Controls \\
\hline

\multicolumn{11}{|l|}{\textit{\textbf{Libraries\textsuperscript{$\dagger$}}}}\\ \hline

\hfill diffprivlib~\cite{IBM_repo} &  I.i, ii \cmark   & II.i \cmark & III.ii, iii \cmark  & \xmark & N/A  & VI.i \cmark & \xmark & \cmark & N/A & N/A \\
\hline

\hfill Google DP~\cite{Google_repo} &  I.i \cmark  & II.ii \cmark & \cmark &  \xmark & N/A  & VI.i \cmark & \xmark & \cmark & N/A &  N/A \\
\hline

\hfill Opacus~\cite{Opacus} & I.ii \cmark  & \xmark & III.ii \cmark &  \xmark & N/A  & VI.i \cmark & \xmark & \cmark & N/A  & N/A \\
\hline

\hfill OpenDP~\cite{SmartNoise_repo} &  I.i \cmark  & II.iii \cmark & III.ii, iii \cmark  & \xmark& N/A  & VI.i, ii \cmark & \cmark & \cmark & N/A  & N/A \\
\hline

\hfill TF Privacy~\cite{tensor_flow_priv} &  I.ii \cmark  & \xmark & \xmark & \xmark & N/A & VI.i \cmark  & \xmark & \cmark  & N/A & N/A \\
\hline

\multicolumn{11}{|l|}{\textit{\textbf{Frameworks\textsuperscript{$\dagger$}}}}\\ \hline

\hfill Chorus~\cite{Chorus_repo} & I.i, iii \cmark & \xmark & \xmark &  \xmark&  N/A & VI.i \cmark &  \xmark & \cmark & \cmark & N/A \\
\hline

\hfill PipelineDP~\cite{pipelineDP}  &  I.i \cmark  & \xmark & III.ii, iii \cmark & \xmark & N/A  &  VI.i \cmark & \xmark &  \cmark  & \xmark &  N/A \\ \hline

\hfill P.~on~Beam~\cite{priv_on_beam}  &  I.i \cmark  & II.ii~\cmark &  \cmark  & \xmark & N/A  &  VI.i \cmark & \xmark  & \cmark   & \xmark & N/A \\ \hline

\hfill Tumult\,Analy.\cite{Tumult_Analytics} & I.i, iii \cmark  & \xmark & \cmark & \xmark & N/A & VI.i, ii \cmark  & \xmark & \cmark  & N/A & N/A \\
\hline

\hfill ZetaSQL~\cite{ZetaSQL} & I.i, iii \cmark & II.ii~\cmark  & \cmark & \xmark& N/A  & \xmark  & \xmark & \cmark  & \xmark & N/A \\ \hline

\multicolumn{11}{|l|}{\textit{\textbf{Systems}}}\\ \hline

\hfill Airavat~\cite{roy_airavat_nodate} & I.i, ii \cmark & \xmark & III.iv \cmark  &  \xmark  &  \xmark  & VI.i, ii \cmark & \xmark & \cmark  & \xmark & \cmark  \\
\hline

\hfill DJoin~\cite{narayan_djoin_nodate} & I.i, iii \cmark & \xmark  & III.ii,~iv,~v~\cmark  & \xmark & \xmark   & VI.i, ii \cmark  & \cmark & \cmark  & \xmark & \xmark \\
\hline

\multicolumn{11}{l}{\textsuperscript{$\dagger$}{\footnotesize Libraries' and frameworks' (III) Security scope is limited to three sub-desiderata (i), (ii), and (iii).}}\\ 

\end{tabular} 

\vspace{-2mm}
\begin{tabular}{|M{3.5cm}|M{1.5cm}|M{1.5cm}|M{1.5cm}|M{1.5cm}|M{1.5cm}|M{1cm}| }

\multicolumn{7}{l}{} \\
\multicolumn{7}{c}{\textit{Table 1B: User interfaces for differential privacy analytics (cf. adapted~\cite{nanayakkara_visualizing_2022}).}} \\
\hline

& \textbf{(V.i)} &  \textbf{(V.ii)} & \textbf{(V.iii)} & \textbf{(V.iv)} & \textbf{(V.v)} & \textbf{(V.vi)} \\ \cline{2-7} 
\centering  \textbf{User Interface/Desiderata}  & Dataset Exploration & Accuracy Visualization & Risk Visualization &  Uncertainty Visualization & Statistical Inference &  Budget Splitting  \\
\hline

\hfill Bittner~et.~al~\cite{bittner} & \xmark & \cmark  & \xmark  & \xmark & \xmark   & \xmark  \\
\hline

\hfill DPcomp~\cite{DPcomp} & \cmark & \cmark  & \xmark  & \xmark & \xmark   & \xmark  \\
\hline

\hfill DPP~\cite{DPP} & \xmark & \cmark  & \cmark  & \xmark & \xmark   & \xmark  \\
\hline

\hfill Overlook~\cite{overlook_DP} & \cmark  & \cmark  & \xmark  & \cmark & \xmark   & \xmark  \\
\hline

\hfill PSI~($\Psi$)~\cite{psi} & \cmark  & \xmark  & \xmark  & \xmark & \xmark   & \cmark  \\
\hline

\hfill ViP~\cite{nanayakkara_visualizing_2022} & \cmark & \cmark  & \cmark  & \cmark & \cmark   & \cmark  \\
\hline

\end{tabular}
%\vspace{-3mm}

\caption{Mapping between open-source tools and user interfaces and the key system desiderata. 
Legend: \cmark = functionality fully available; \xmark = limited functionality or not available; N/A = not applicable; P. = Privacy; DP = Differential P.; TF = TensorFlow; I.i = Enables aggregation queries; I.ii = Enables machine learning; I.iii = Enables query clauses (e.g., {\tt JOIN}); II.i = Query semantic consistency; II.ii = DP sensitivity calculation; II.iii = Privacy parameter search; III.i = DP correctness verification; III.ii = Cryptographically secure pseudo-random number generation; III.iii = Protection against floating-point vulnerability; III.iv = Block data visibility; III.v = Block arbitrary code; VI.i = Budget accountant; VI.ii = Query blocker.\newline} 
\label{tab:mapping_tools_requirements}
%\vspace{-12pt}
\end{table*}

In secondary use cases, an alternative to \emph{syntactic} anonymization (see section~\ref{subsec:Anonymizing_Data}) for sharing data is an inherently private analysis, i.e., the analysis satisfies a \emph{semantic} privacy definition such as DP~\cite{syntactic_semantic_priv}, which uniquely provides a measure of privacy ($\varepsilon$).
With DP, organizations do not necessarily need to use potentially vulnerable syntactic techniques (e.g., rounding or truncation) because the analysis itself already enhances individuals' privacy.
Based on the (i) interviewees' description of their analytics workflows and systems, (ii) the authors' knowledge in the domain of privacy, and (iii) the feedback provided by additional privacy practitioners and researchers who work closely with/in our lab, we propose $10$ key desiderata.
The desiderata correspond to a system that enables differentially-private analyses in the central model and focuses on dataset exploration and fulfilling use cases requiring aggregations (see use cases in Q12).
These use cases often rely on SQL-like queries such as counts, averages, etc.
Additionally, we inspired some of the characteristics of the key desiderata related to (III) \textit{Security} and (V) \textit{Visualization} from Kifer.~et.~al~\cite{meta_DP} and Nanayakkara.~et.~al~\cite{nanayakkara_visualizing_2022}, respectively.
\smallskip 

 %\item \textit{Re-writes aggregation queries to fulfill differential privacy}. 
    %The system bestows differential privacy to the query by adding calibrated noise and, after the database engine ingests and executes the re-written query, the system returns the differentially private result to the analyst.
\noindent \textbf{(I)} \textit{Differentially private analytics}. 
The system bestows DP to a learning function (e.g., a query or an ML algorithm) by adding calibrated noise to the deterministic outputs (or by other means). 
The system supports the (i) aggregation queries: {\tt COUNT}, {\tt MAX}, {\tt MIN}, {\tt AVG}, {\tt VAR}, and {\tt SUM}, (ii) provides a complementary ML feature, and stores executed queries for future retrieval.
The queries (iii) allow for {\tt WHERE}, {\tt GROUP\,BY}, and {\tt JOIN} clauses.
%Altogether, the system offers an interface to execute (at least) aggregation queries that guarantee DP for exploring a dataset or fulfilling use cases (if possible).
\newline 
\noindent \textbf{(II)} \textit{Usability}. 
The system provides logic to preserve (i) the semantic consistency of queries (e.g., variance > $0$) and across overlapping domains (e.g., the sum of noisy element counts is not larger than the noisy total). Moreover, the system presents the option to (ii) estimate the sensitivity of a query without user input, and (iii) recommends or sets privacy parameters automatically depending on the dataset and query.
\newline 
\noindent \textbf{(III)} \textit{Security}. 
The system (i) provides a stochastic tester or other functions to automatically verify whether the algorithm fulfills DP, (ii) employs cryptographically secure pseudo-random number generation with careful seed management, (iii) generates noise impervious to floating-point vulnerabilities~\cite{mironov_significance_2012, precision_attacks}. Furthermore, the system (iv) blocks the user from ``\textit{seeing}'' the data, i.e., while analysts can execute queries, they cannot download or visually inspect the dataset, (v) does not allow to execute arbitrary code, (vi) executes heuristic optimizers only at post-processing, and (vii) protects against timing attacks~\cite{meta_DP}. A libraries' and frameworks' scope limits to fulfilling (i), (ii), and (iii).
\newline 
\noindent \textbf{(IV)} \textit{Synthetic data generation}.
When the goal is to develop an application or explore whether an ML model is suitable for a task, the system produces synthetic data. After testing, the analyst can proceed with the real data (without ``\textit{seeing}'' it).
Synthetic data generation could rely on simple techniques (e.g., sampling from a normal distribution with the same mean and standard deviation as the target attribute), ML~\cite{DP_synth_bench, synthetic_data_generation, synthetic_data_permute_2}, or combining DP with either.
If the analyst is only interested in the data schema, the system produces dummy data, preserving only the schema and data types.
\newline 
\noindent \textbf{(V)} \textit{Visualization}. The system presents a dashboard depicting interactive plots (e.g., histograms) relying on DP queries for quick and intuitive (i) dataset exploration. 
Additionally, the dashboard visualizes an analysis' expected (ii) accuracy and (iii) disclosure risk, (iv) uncertainty (i.e., a measure of how the same mechanism can produce different outputs with the same input arguments), (v) statistical inference (i.e., privacy parameter estimation with confidence intervals), and (vi) budget splitting (i.e., help in splitting the privacy budget across queries)~\cite{nanayakkara_visualizing_2022}.
\newline 
%\noindent \textbf{(V)} \textit{Provides acceptable execution time}. Bestowing differential privacy to the queries should not become a bottleneck.
%\newline 
\noindent \textbf{(VI)} \textit{Privacy budget}. The system (i) tracks the budget spent ($\varepsilon$ ``odometer''), (ii) blocks further queries if analysts exhaust their budget, and (iii) accommodates the budget for growing datasets. (iv) It should enable data stewards to specify budgets for teams, individual analysts, and use cases depending on the data's sensitivity. % (see privacy budget concept in section~\ref{sec:differential_Privacy}).
\newline 
\noindent \textbf{(VII)} \textit{Accuracy adjustment}. The system allows the user to propose a desired accuracy level. 
Alternatively, after the query execution, the system provides either information about the noise scales (without additional budget) or a confidence interval (spending budget)~\cite{Vesga2020APF}.
\newline 
\noindent \textbf{(VIII)} \textit{Query sensitivity}. The system enables a  practitioner, e.g., a data analyst or steward, to input the attributes' bounds (maximum and minimum values) as function parameters or in the dataset schema so that the system calculates the sensitivity of the query to calibrate the noise. 
\newline 
\noindent \textbf{(IX)} \textit{Privacy-sensitive data annotation}. The system enables data stewards to allowlist attributes based on teams, roles, and use cases. The system automatically obfuscates attributes not contained in the allowlist.
\newline 
\noindent \textbf{(X)} \textit{Authentication and access controls}. The system easily integrates with existing authentication and access control services and enables data stewards to define their access policies. 
\vspace{-4mm}

\begin{comment}
Lastly, we argue that if the answer to an aggregation analysis is a single output (e.g., \textit{what is the most frequented location?}) not used in combination with others, the system should output the exact answer without revealing any user information (e.g., not revealing the aggregate frequencies of the users).
Nonetheless, we encourage researchers to study its privacy risks.
\end{comment}

\subsection{Gaps in Differential Privacy Practice}
\label{subsec:gap_in_DP}

Despite available open-source tooling, one company found it hard to find external partners that could bring DP into practice in their internal analysis workflow.
Furthermore, another company stated after exploring the use of DP that, while it seemed helpful, ``\textit{[Deploying differential privacy] was more expensive than doing nothing}."
Instead, the department decided to upload syntactically anonymized data to a highly secured system, with limitations on access time, downloads, number of analysts, and audit logs.
We kindly argue that their over-statement was due to the intangible costs of the dataset request processes and the lack of integrability of current DP tooling, which makes deployment a complex endeavor.

Overall, our findings indicate a gap between the theory and practice of DP.
Working towards bridging the gap, we qualitatively mapped in Table~\ref{tab:mapping_tools_requirements} our key system desiderata with DP tools to highlight areas of future work for the privacy community.
We selected the tools from the related work in section~\ref{sec:Related_Work} that offer open-source implementations for the central model of DP (see tool descriptions in Appendix~\ref{app_sec:Open-source_tools_descriptions}).
We must highlight that some of these tools are \emph{libraries} (provide specific functions) or \emph{frameworks} (abstractions used to build specific applications) and, thus, lack functionalities that a \emph{system} (end-to-end application) like Airavat~\cite{roy_airavat_nodate} could provide, such as (III.iv) \textit{Blocking the visibility of data} or (X) \textit{Authentication and access controls}. Note that libraries and frameworks assume analysts have data access.
Additionally, we regard \emph{user interfaces} (systems focused on visualizations and providing analytics metadata) as a set of tools that should fulfill key desiderata specific to (V) \textit{Visualization}.
Accordingly, we assign each open-source software to its category in Tables~\ref{tab:mapping_tools_requirements}A and~B for an appropriate comparison.

We must highlight that the mapping of Table~\ref{tab:mapping_tools_requirements} provides high-level guidance, as there are (out-of-scope) nuances Table~\ref{tab:mapping_tools_requirements} does not capture.
For example, user interfaces such as Bittner~et.~al~\cite{bittner} and DPP~\cite{DPP} in Table~\ref{tab:mapping_tools_requirements}B provide exploratory results for using DP ML and for disclosure risk, respectively; however, they do not help understanding the dataset, which is a critical requirement for data analysts.
Regarding the tools in Table~\ref{tab:mapping_tools_requirements}A, diffprivlib~\cite{IBM_repo} offers multiple ML models (PCA, Naive Bayes, liner and logistic regression, k-means) while others focus on deep learning (Opacus \cite{Opacus} and TensorFlow (TF) Privacy \cite{tensor_flow_priv}) or MapReduce functionality (Airavat). Additionally, the frameworks are designed for large-scale datasets.
We note that Google DP~\cite{Google_repo} provides the building blocks for ZetaSQL~\cite{ZetaSQL} and Privacy on Beam~\cite{priv_on_beam} (and PipelineDP~\cite{pipelineDP}), which add more functionality for considering datasets with multiple individual's contributions. Lastly, most tools provide only an ``odometer'' for privacy budgeting, while a few block new queries if the budget is spent (e.g., OpenDP~\cite{SmartNoise_repo} and Tumult Analytics~\cite{Tumult_Analytics}), and Google DP offers functionality to distribute budget across different DP mechanisms~\cite{PDL, concentrated_DP}. 
None, however, account for growing datasets, which is a challenge recently tackled in~\cite{priv_account_pro}. 
One may find more of these nuances in~\cite{DP_benchmark}.

Based on the non-availability or limited implementations of some desiderata in Table~\ref{tab:mapping_tools_requirements}, we conclude that \emph{differential privacy tool designers can learn from one another}, \textit{no tool outperforms the rest in every aspect}, and, most importantly, that \emph{bridging the gap is primarily an engineering problem}.
Subsequently, we identify the major gaps in differential privacy practice:

\noindent\noindent \textbf{Gap 1:} (II) \textit{Usability}. 
While semantic consistency is sometimes desirable for analysts, it can also introduce more error/bias in some scenarios. Only diffprivlib implements mechanisms to fulfill DP and consistency for specific queries (e.g., variance > $0$), whereas Google DP or Tumult Analytics only truncate values~in post-processing. Furthermore, only Google DP can calculate the query sensitivity in a privacy-enhancing manner without any user input, which is necessary when an analyst lacks domain knowledge of the application (i.e., input bounds).
Thus, none of the tools in Table \ref{tab:mapping_tools_requirements} completely fulfill the usability desiderata.
\textit{Guidance}: \cite{Google_repo, IBM_repo, DP_consistency_1, DP_consistency_2, DP_consistency_3, Linkedin_usenix, rogers_linkedins_2021}
\newline
\noindent\noindent \textbf{Gap 2:} (III) \textit{Security}.
The tools do not provide many security features individually. 
For example, most lack stochastic testers to verify that an analysis fulfills DP, and none implement protections against time-attacks~\cite{meta_DP}.
Wrt to secure random number generation: TF Privacy inherits the insecure RNG from TF~\cite{TF_P_rand, TF_rand} and Airavat employs the insecure utility {\tt java.util.Random}~\cite{airavat_vulnerability} in contrast to DJoin, which relies on FairplayMP~\cite{djoin_not_vul, fairplay}.
Moreover, we encourage developers of Opacus and TF Privacy to include floating-point protections in their deep learning models or rely on discrete noise distributions~\cite{geometric_mechanism, discrete_Gaussian} like Tumult Analytics. Moreover, most tools should tackle their precision-based attack vulnerability~\cite{precision_attacks}.
Lastly, we highlight some of the good practices Kifer~et.~al~\cite{meta_DP} proposed, namely, open-sourcing systems (the community can check for vulnerabilities) and performing code audits and unit tests to ensure correctness in DP, privacy accounting, and noise sampling. \textit{Guidance}:~\cite{Google_repo, meta_DP, Google_Plum}
\newline 
\noindent\noindent \textbf{Gap 3:} (IV) \textit{Synthetic data generation} (SDG). Similarly to tools offering DP ML~\cite{IBM_repo, tensor_flow_priv, Opacus, roy_airavat_nodate}, we suggest developers package and include DP SDG logic. \textit{Guidance}:~\cite{DP_synth_bench, synth_DP_1, synth_DP_2, synth_DP_FL, synthetic_data_generation, synthetic_data_permute_2, chen2020gs, torkzadehmahani2019dp}.
\newline
\noindent \textbf{Gap 4:} (V) \textit{Visualization}. While there is enough research on user interfaces, the most popular frameworks and libraries do not adopt them. We suggest packaging available DP user interfaces for patching analytics tools. \textit{Guidance}:~\cite{nanayakkara_visualizing_2022, overlook_DP}. 
\newline
\noindent \textbf{Gap 5:} (VI) \textit{Privacy budget}. A surprisingly high number of tools implement privacy ``odometers'' without a logic to block queries after exceeding the budget. \textit{Guidance}:~\cite{priv_on_beam, narayan_djoin_nodate, roy_airavat_nodate}.
\newline
\noindent \textbf{Gap 6:} (VII) \textit{Accuracy adjustment}. While most user interfaces provide some form of accuracy calculation and visualization, many other tools overlook such feature. \textit{Guidance}:~\cite{Vesga2020APF, nanayakkara_visualizing_2022}.
\newline
\noindent \textbf{Gap 7:} (IX \& X) \textit{Functionality for data stewards}. Only a few tools enable data stewards and owners to (IX) annotate sensitive data and (X) define and enforce access controls. Developers do not need to reinvent the wheel, as they adopt current best practices from popular cloud platforms~\cite{AWS, azure, google_cloud}. \textit{Guidance}:~\cite{Chorus_repo, DPP, roy_airavat_nodate}.

Given that most functional requirements are available in components across tools, we conclude that \textit{engineering efforts are within striking distance}.
To complement these building blocks, we offer an early stage, high-level system design blueprint in Appendix~\ref{app_sec:system_design}. Our design aims to spark interest in practitioners to develop holistic privacy-enhancing analytics tooling that follows the identified key system desiderata.

\vspace{-1mm}
\section{Further Challenges}
\label{sec:discussion}

Beyond the engineering and organizational challenges discussed in the previous sections, there exist other critical technical challenges in DP. 
In combination: Managing privacy budgets on large-scale user data streams \cite{priv_account_pro} with unknown domains and user contributions on multiple records \cite{Google_Plum} across different systems while adapting the noise level as the budget diminishes.
Furthermore, fitting a mathematical model to such a system's semantics and verifying DP fulfillment with, e.g., unit tests, poses additional difficulties~\cite{meta_DP}.
Additionally, DP might not be \textit{fair}~\cite{kenny_use_2021} in some use cases where a DP calculation determines a critical outcome, e.g., a user's financial support in an underwriting model (see challenges in Appendix~\ref{app_sec:challenges_dp}).

Our work highlights the challenges blocking the broader adoption of DP in organizations' workflows. Dwork~et~al.~\cite{dwork_differential_2019} partly studied these challenges by interviewing DP experts, while our study brings non-experts considerations into the discussion.
Dwork~et.~al distilled $4$ main challenges from their interviews (section~3.6~\cite{dwork_differential_2019}), which overlap with a few of our findings:
(i) \textit{Part of the challenges deploying DP were design based}. In section~\ref{sec:Query_Re_Writer}, we highlight that current DP tools still require engineering effort to be easily deployable in organizations.
(ii) \textit{DP deployment complexity is also institutionally based}. A common theme of the interviewed companies was their intricate networks of stakeholders and processes, which hamper goal alignment and technology deployments.
(iii) \textit{There was no consistency in DP approaches across institutions, indicating a need for shared learning}. One of our conclusions in section~\ref{sec:Query_Re_Writer} signals that tool designers can learn from one another.
(iv) \textit{Transparency and testable statements about privacy can benefit companies in the regulatory landscape}. Similarly, in section~\ref{sec:Query_Re_Writer}, we advocate for transparency in system designs and moving towards DP-centered systems and away from syntactic privacy definitions that only guarantee \textit{factual anonymity}.

\noindent \textbf{Future work.} We suggest privacy practitioners fill the gaps highlighted in section~\ref{sec:Query_Re_Writer} and tackle the challenges of Appendix~\ref{app_sec:challenges_dp}. Moreover, specifically for privacy researchers, we encourage (i) improving guidance on selecting $\varepsilon$~\cite{dwork_differential_2019, DPP} and (ii) studying and communicating to non-experts how mechanism designs affect utility. For example, studying how output consistency can imbue bias~\cite{DP_consistency_1, DP_consistency_2, DP_consistency_3} or floating-point protection may provide less utility.
As new DP deployments increasingly resort to more complicated algorithms~\cite{range_queries_DP}, we suggest (iii) studying the unpredictable artifacts these algorithms may introduce (e.g., in the 2020 US Census~\cite{census_issues_DP}). Lastly, we encourage improving current proposals of differentially private (iv) ML and (v) synthetic data generation.

\vspace{-2pt}
\section{Conclusion}
\label{sec:Conclusion}

We conclude that DP can improve the work of data scientists across industries by enabling sensitive data exploration across silos, potentially shortening data access times by relaxing the adversity of data request processes, and can fulfill some types of use cases. Furthermore, analysts meaningfully and frequently employed analyses amenable to DP and, on average, would feel comfortable with a $98\%$ of accuracy. Therefore, we suggest companies focus on privacy-enhancing analysis to harvest these benefits, not mainly on security. Moreover, we regard enabling analysts to work without ``\textit{seeing}'' data and providing analysis accuracy expectation as critical, multifaceted challenges for the research community to solve. We also highlight that current open-source tools do not facilitate easy deployments, a problem requiring engineering effort within striking distance. Consequently, we encourage the community of privacy practitioners to tackle this engineering problem and ease deployments by enabling interactive dashboards, accuracy expectation measurements, improving usability and security, and integration of data annotation and access control capabilities, for ultimately bridging the identified gap between theory and practice.

%Given these road-blocks, the interviewed companies have not adopted differential privacy, despite being leaders in their respective industries.

\begin{comment}

Some countries do not care, but they push the GDPR standard, whereas in Germany they are on the learning side.

Despite the federation, analyst must interview with stewards.

tech companies like Google or Apple where they can pull data from devices with the the same OS, data formatting, ... have a huge advantage

About privacy for the premium product - build trust and differentiate your brand. Abiding the law - prevent risk.

\end{comment}

\bibliographystyle{ACM-Reference-Format}
\bibliography{999_REFS}

\appendix

\section{Interviewed Companies Overview}
\label{app_sec:interviewed_companies}

Table~\ref{tab:interviewed_companies} presents a summary of the characteristics of the interviewed companies.
The companies belong to a diverse set of industries, predominantly SW development, and $4$ of the $9$ companies are significantly large, with over $100,000$ employees.
$5$ companies are under the jurisdiction of the EU with regulations such as the General Data Protection Regulation~\cite{GDPR_EU2016}, and $4$ companies operate under US law, e.g., the California Consumer Privacy Act~\cite{ccpa2018}.

\begin{table} [htpb!]
\centering
\small
\begin{tabular}{|M{4.6cm}|M{1.35cm}|M{1.15cm}|}
\hline

 \textbf{Industry (focus)} & 
\textbf{Size \break (employees)} & \textbf{Team's Location}
\\
\hline 

\multicolumn{3}{|l|}{\textbf{Team operates internationally}} 
\\ \hline
 Automotive (car manufacturer) & $>100,000$  & Germany 
\\\hline
 Insurance (health) & $>100,000$  & Germany 
\\\hline
 SW dev. (data processing) & $<2,000$  & Germany 
\\\hline
 SW dev. (subscription newsletters) & $<2,000$  & USA
\\\hline
\multicolumn{3}{|l|}{\textbf{Team operates nationally}} 
\\ \hline
 Consultancy (banking and big pharma) & $>100,000$ & Spain
\\\hline
 Entertainment (finance) & $<2,000$ & USA 
\\\hline
 SW dev. (business operations) & $>100,000$   & Germany
\\\hline
 SW dev. (data processing) & $<2,000$  & USA
\\\hline
 SW dev. (smart sound system) & $<2,000$  & USA 
\\\hline

\end{tabular}
\caption{Overview of the interviewed companies. Legend: SW dev. = Software development} 
\label{tab:interviewed_companies}
%\vspace{-12pt}
\end{table}

\section{Interview Questionnaire for Data Stewards}
\label{app_sec:Interview_questionnaire_stewards}

\begin{enumerate}
    \item[]  \textbf{RQ1:} \textit{What is the context of privacy protection in the targeted organization?}
    \begin{enumerate}
      \item[] \textbf{Q1:} \textit{What is the institution's motivation for privacy protection?}
      \item[] \textbf{Q2:} \textit{What are your privacy concerns when an analyst has full dataset access? } 
      \item[] \textbf{Q3:} \textit{At what level of data granularity are you protecting and measuring privacy?}
      \item[] \textbf{Q4:} \textit{What could be improved in the dataset request process?}
      \item[] \textbf{Q5:} \textit{What are your typical questions for the current interview-based full dataset access authorization? }
      \item[] \textbf{Q6:} \textit{Instead of the interview process, would you be capable to run a program provided by the analyst s.t. the analysis is carried out without the analyst ever ``seeing'' the dataset? }
    \end{enumerate}
\end{enumerate}

\section{Interview Questionnaire for Data Analysts}
\label{app_sec:Interview_questionnaire_analysts}

As we interviewed non-experts in differential privacy, we minimized the number of questions that contained the words or required knowledge of ``\textit{differential privacy}''.
We kept a few because we aimed to assess whether systems offering differential privacy functionality could be valuable to analysts.
First, we briefly explained differential privacy in a simplified manner: ``\textit{Differential privacy is a technique that adds noise to analytics results so that one cannot reverse engineer the outputs to a specific person.}''
Additionally, if we perceived the interviewees were disoriented with Q18 or Q19, we explained that the hypothetical system would be the same as the one they used every day, the only difference being that the results would slightly differ from the deterministic outputs.
Picturing the system they used daily was very helpful for imagining one where the outputs are noisy.
Furthermore, we carefully parsed their answers to assess whether they understood the concept or its integration into their system. If they did not, we kindly repeated the procedure above.

\begin{enumerate}
  \item[] \textbf{RQ2:} \textit{Could differential privacy tackle the privacy-related pain points of an analysis workflow in an organization?} 
  \begin{enumerate}
      \item[] \textbf{Q7:} \textit{What is your workflow to analyze data? }
      \item[] \textbf{Q8:} \textit{Why do you need full dataset access?}
      \item[] \textbf{Q9:} \textit{How often do you request full dataset access? How long does it usually take?}
      \item[] \textbf{Q10:} \textit{What do you think about the process to request full dataset access in your organization?}
      \item[] \textbf{Q11:} \textit{What features do you think are missing in your organization's data analysis workflow?}
  \end{enumerate} \smallskip

  \item[] \textbf{RQ3:} \textit{When does differential privacy impede an analysis?}
  \begin{enumerate}
      \item[] \textbf{Q12:} \textit{In which analytics use cases have you been involved?}
      \item[] \textbf{Q13:} \textit{Is SQL-meaningful for your work? How many SQL-like queries do you make weekly?}
      \item[] \textbf{Q14:} \textit{How often do you need machine learning to fulfill your analysis in contrast to using SQL?}
      \item[] \textbf{Q15:} \textit{What are your most used machine learning models?} 
      \item[] \textbf{Q16:} \textit{If you were to use differential privacy to fulfill your analysis, when and how much accuracy would you be willing to forgo?} 
  \end{enumerate} \smallskip
 
  \item[] \textbf{RQ4:} \textit{How would differential privacy affect the workflow of an analyst?} 
  \begin{enumerate}
      \item[] \textbf{Q17:} \textit{How much would the noise affect your analysis?}
      \item[] \textbf{Q18:} \textit{Would you find it helpful to execute differentially private SQL queries to explore and fully analyse datasets without the standard permissions?}
      \item[]  \textbf{Q19:} \textit{Only based on the information extracted from a dataset exploration with differential privacy, could you write a script to fulfill your analysis goal?}
      \item[] \textbf{Q20:} \textit{What are the minimum properties for you as an analyst such that you are confident to write an analysis script without full dataset access?}
      \item[] \textbf{Q21:} \textit{Would you find it helpful to use a dynamic dashboard that visualizes dataset information with differential privacy?}
  \end{enumerate} \smallskip

  \item[] \textbf{RQ5:} \textit{Can differential privacy be applied to the frequent SQL-like queries analysts execute?}
  \begin{enumerate}
      \item[] \textbf{Q22:} \textit{What are your top SQL-like queries before you have full dataset access?}
      \item[] \textbf{Q23:} \textit{What are your top SQL-like queries after you have full dataset access?}
      \item[] \textbf{Q24:} \textit{What is the ratio between aggregation queries and queries to retrieve items?}
  \end{enumerate}
\end{enumerate}

\section{Frequent Queries}
\label{app_sec:frequent_queries}

In Table~\ref{tab:frequent_queries}, we include the most frequent SQL queries recorded during the interviews with the data analysts before and after accessing a dataset.
Note that not all analysts were allowed to explore datasets and a few did not employ SQL for data preparation or analytics; instead, they resorted to Python scripts for statistical analysis, ML, and visualization or tools such as Tableau~\cite{Tableau}, Knime~\cite{Knime}, or proprietary SAP data management software.
For exploring the dataset prior to access, $14$ analysts resourced to {\tt SELECT~*} to get a ``\textit{feeling}'' for the data.
Furthermore, {\tt COUNT} and {\tt DISTINCT}, and {\tt WHERE} and {\tt GROUP BY} were the most frequently used functions and clauses, respectively.

\begin{table} [htpb!]
\centering
\small
\begin{tabular}{|M{2cm}|M{1.6cm}|M{1.6cm}| }
\hline

\textbf{Query} & 
\textbf{Freq. before} & \textbf{Freq. after }
\\

& 
\textbf{access} & \textbf{access}
\\\hline

\multicolumn{3}{|l|}{\textbf{Function}} 
\\ \hline

\hfill {\tt COUNT}  & $7$  & $7$
\\\hline

\hfill {\tt DISTINCT}  & $6$  & $4$
\\\hline

\hfill {\tt MAX}  & $4$  & $5$
\\\hline

\hfill {\tt MIN}  & $4$  & $5$
\\\hline

\hfill {\tt AVG}  & $4$  & $4$
\\\hline

\hfill {\tt VAR}  & $2$  & $3$
\\\hline

\multicolumn{3}{|l|}{\textbf{Statement}} 
\\ \hline

\hfill {\tt SELECT~* LIMIT}  & $14$  & $12$ 
\\\hline

\multicolumn{3}{|l|}{\textbf{Clause}} 
\\ \hline

\hfill {\tt WHERE}  & $13$  & $10$ 
\\\hline

\hfill {\tt GROUP BY}  & $12$  & $9$ 
\\\hline

\hfill {\tt JOIN}  & $2$ & $8$ 
\\\hline

\end{tabular}

\caption{Most frequent queries before (data exploration) and after (data preparation/analysis) data access. Legend: Freq. = Frequency (i.e., number of analysts who used such query).} 
\label{tab:frequent_queries}
%\vspace{-12pt}
\end{table}

\section{Frequently Asked Questions from Data Stewards to Analysts}
\label{app_sec:stewards_questions}

We compiled the most frequently asked questions data stewards make to data analysts during the data access request process.

\begin{itemize}
  \item Could you describe in detail the analytics use case? 
  \item Is the use case approved by the corresponding internal stakeholders?   
  \item Why is the dataset needed?
  \item Is the dataset adequate regarding quality, volume, and use case?  
  \item Could you reach the goal without dataset access?
  \item Is the entire dataset needed or only a set of attributes?
  \item Is the dataset already available, or must a data engineer create a new dataset?
  \item Is the dataset classified as very sensitive? If affirmative, additional access control measures and monitoring must be defined in detail.
\end{itemize}

\section{Open-source tools descriptions}
\label{app_sec:Open-source_tools_descriptions}

We provide a quick description of each of the selected open-source tools mapped to the key system desiderata in section~\ref{sec:Query_Re_Writer} appearing in Table~\ref{tab:mapping_tools_requirements}. \smallskip

\noindent \textit{\textbf{Libraries}} 
\newline 
\noindent \textbf{diffprivlib}: IBM researchers developed a general-purpose Python library to execute differentially private aggregation queries and machine learning in the context of data science (namely Notebooks)~\cite{IBM_repo}.
\newline 
\noindent \textbf{Google DP}: Google researchers developed a library in multiple languages (C\Plus\Plus, Go, and Java) that an expert may use to build new applications supporting differential privacy~\cite{Google_repo}.
\newline 
\noindent \textbf{Opacus}: Meta researchers developed a library dedicated to training machine learning models offered by PyTorch in a differentially private manner~\cite{Opacus}.
\newline 
\noindent \textbf{OpenDP}: Harvard implemented a flexible architecture for differentially private analysis, consisting of a (pluggable) runtime in Rust wrapped around a Python API, in addition to a ``validator'' that calculates parameters such as the sensitivity of a query.~\cite{SmartNoise_repo}.
\newline 
\noindent \textbf{TensorFlow Privacy}: Google researchers developed a library that includes TensorFlow differentially-private optimizers for training machine learning models~\cite{tensor_flow_priv}. \newline

\vspace{2mm}
\noindent \textit{\textbf{Frameworks}} 
\newline 
\noindent \textbf{Chorus}: Johnson et al.'s~\cite{Chorus_repo, Uber_DP} wrote a framework in Scala that works in cooperation with existing infrastructure (a SQL database) to explore the use of differentially private SQL queries at scale.
\newline 
\noindent \textbf{PipelineDP} (experimental): OpenMined, in collaboration with Google, propose a framework to execute differentially private aggregations in large-scale datasets using batch processing systems (Apache Spark and Apache Beam)~\cite{pipelineDP}.
\newline 
\noindent \textbf{Privacy on Beam}: Similarly to PipelineDP, Privacy on Beam~\cite{priv_on_beam} proposes a solution based on Apache Beam and Google DP~\cite{Google_repo} for executing differentially private analytics at scale. 
\newline 
\noindent \textbf{Tumult Analytics}: Tumult Labs provides a Python library built atop a framework similar to OpenDP for computing aggregate statistics over tabular data at scale~\cite{Tumult_Analytics}.
\newline 
\noindent \textbf{ZetaSQL}: Google researchers wrote a framework for SQL that defines a language, a parser, and an analyzer meant to work with an existing database engine~\cite{ZetaSQL}.
    
\vspace{2mm}
\noindent \textit{\textbf{Systems:}}
\newline 
\noindent \textbf{Airavat}: Roy et al.~~\cite{roy_airavat_nodate} designed a MapReduce-base system written in Java for distributed computations on sensitive data that integrates differential privacy and access control with policies defined by data owners/stewards.
\newline 
\noindent \textbf{DJoin}: Narayan et al.~\cite{narayan_djoin_nodate} built a system capable of processing a wide range of differentially private SQL queries across datasets from different organizations and leverages homomorphic primitives to hide inputs. \smallskip

\noindent \textit{\textbf{User Interfaces:}} 
\newline 
\noindent \textbf{Bittner~et.~al~\cite{bittner}}: With a focus on ML, Bittner~et.~al aim to help researchers decide which algorithm to use by offering an interface that quantifies the disclosure risk of different algorithms.
\newline 
\noindent \textbf{DPcomp}: A web-based system enabling researchers to assess the utility of differentially private algorithms and understand their respective incurred error~\cite{DPcomp}.
\newline 
\noindent \textbf{DPP}: This user interface specifically helps data owners to set the noise level per the disclosure risk of an attribute. The underlying mechanism relies on a novel parameter selection procedure for differential privacy~\cite{DPP}.
\newline 
\noindent \textbf{Overlook}: Thaker~et~al.~\cite{overlook_DP} designed a system for differentially private data exploration that supports counts with an interactive browser-interfacing dashboard (namely visualizing histograms).
\newline 
\noindent \textbf{PSI~($\Psi$)}: Harvard's Privacy Tools Project works on a data sharing interfaces for researchers to explore datasets with differential privacy~\cite{psi}.
\newline 
\noindent \textbf{ViP}: Visualizing Privacy is an interface that provides information about the relationships between utility, $\varepsilon$, and disclosure risk (among others), allowing users to adjust the privacy parameters of their analysis based on visualizations of expected risk and accuracy~\cite{nanayakkara_visualizing_2022}.

\section{Differential Privacy challenges}
\label{app_sec:challenges_dp}

This section enumerates other critical challenges we encourage researchers and system designers to investigate.

\noindent \textbf{(1)} While DP is highly adaptable to use cases (e.g., using the local or central model) and algorithms (e.g., queries or ML), the adaptations are non-trivial and have often led to erroneous implementations~\cite{domingo_ferrer_limits_2021}.
Thus, practitioners should exercise extreme care to ensure the correctness of their DP implementation with the same sentiment as ``\textit{do not write your own crypto}.''

\noindent \textbf{(2)} \emph{Fairness} could be another obstacle to DP adoption, which Harvard researchers also highlighted when referring to the US Census of 2020~\cite{kenny_use_2021}. Specifically, one analyst underlined the topic of fairness when asked about how noise would affect their analysis (Q17).
If analysts add differentially private noise during training underwriting linear regression models, users might be over- or under-funded.
While the company would not incur a loss as the predictions would be ``right'' on average, the effect noise has on their users could impact their brand perception.
    
\noindent \textbf{(3)} Managing \emph{user-level} privacy budgets in user data streams~\cite{priv_account_pro}.

\noindent \textbf{(4)} Tracking the privacy budget across systems and adapting the noise level based on the remaining budget.

\noindent \textbf{(5)} There is a significant difference between the local and central model noise levels.

\noindent \textbf{(6)} Choosing $\varepsilon$ and other privacy parameters~\cite{dwork_differential_2019}.

\noindent \textbf{(7)} Building systems that fulfill DP for (i) large-scale datasets (ii) when users make contributions to multiple records (iii) with unknown domains~\cite{Google_Plum}.  

\noindent \textbf{(8)} Verifying DP compliance of a complex system by proving and fitting a mathematical model to the system's semantics~\cite{meta_DP} and developing unit tests to ensure the system conforms to such model.

% Such an example is another reminder of how crucial it is to understand the implications of analysis and the consequences that individuals could suffer.

\section{System Design}
\label{app_sec:system_design}

Although there are many potential ways to construct privacy-enhancing analytics systems, to show the feasibility of covering all system desiderata presented in section~\ref{sec:Query_Re_Writer}, this section discusses one design to guide practitioners in their development.
The design is in an early stage, and, thus, we cannot discuss the components in detail.
Instead, we sketch the system's primary components, aiming to spark interest in further system development and research in the community.
% Based on the key system desiderata of section~\ref{sec:Query_Re_Writer}, we designed a high-level blueprint of a privacy-enhancing analytics system to guide practitioners in their development further.

\begin{figure*}[!t]
    \centering
    \includegraphics[scale=0.7]{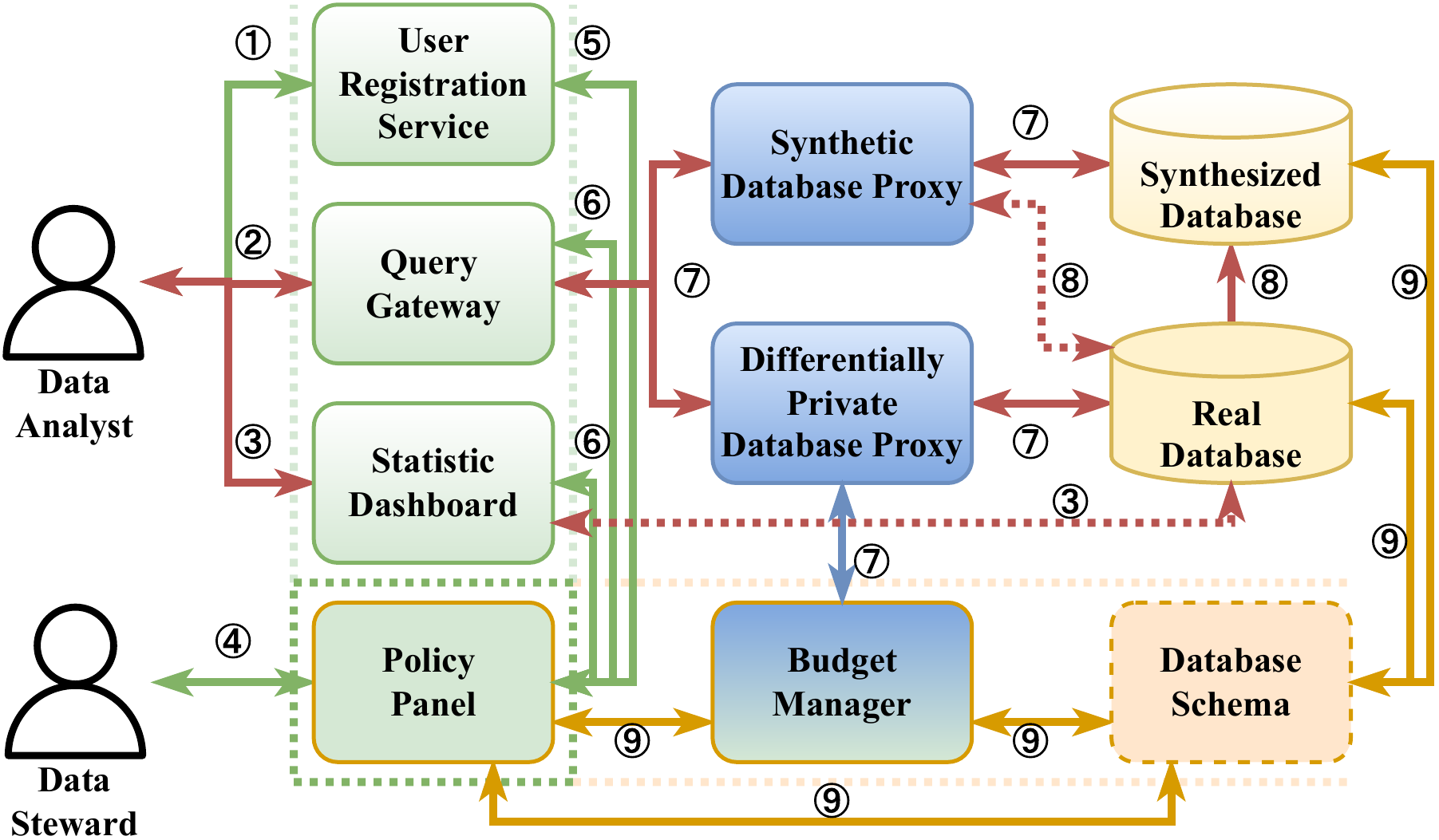}
    \caption{High-level system design blueprint of a privacy-enhancing analytics tool. The components and communication links are described in Appendix~\ref{app_sec:system_design}. Solid lines represent communications between components triggered by all relevant query events, while dashed lines represent communications that happen periodically or only under certain circumstances. We specify those circumstances in the workflow description.}
    \label{fig:system_design}
\end{figure*}

We consider two roles interacting with the system: (i) \emph{data stewards} have the authority to access the original data and the legal background for data management. Stewards can authorize data access inside an organization and ensure compliance. 
(ii) \emph{Data analysts} analyze data to fulfill use cases. Analysts often need to access data by employing SQL aggregation or retrieval and python scripts.
In the current system design, we mainly consider SQL aggregation and retrieval. 
Lastly, we assume that analysts cannot share query results with unauthorized recipients.
In such a setting, we present our high-level system design blueprint in Fig.~\ref{fig:system_design}. 
The components are the following:

\noindent \textbf{Database schema}: The system requires one dedicated component to manage the database schema and ensure its consistency at all places to make sure all components have a consistent view of the processed data format. \newline
\noindent \textbf{Policy panel}: Data stewards create and update the configuration stored in the policy panel to authorize data access from data analysts to satisfy desiderata \textbf{(X)}, annotate data sensitivity to satisfy desiderata \textbf{(VIII)} and \textbf{(IX)}, and ensure compliance. Other system components rely on the configuration in the policy panel to decide whether to proceed with particular requests or queries. \newline
\noindent \textbf{User registration service}: The user registration service component maintains a user system to standardize the onboarding procedure of data analysts to satisfy desiderata \textbf{(X)}; thus, the system can distinguish between different data analysts with different data access requirements and permissions. \newline
\noindent \textbf{Statistic dashboard}: The statistic dashboard is a privacy-enhanced visualization for database statistics, which will help authorized data analysts explore datasets, thus satisfying desiderata \textbf{(V)}. \newline
\noindent \textbf{Query gateway}:  The query gateway reads annotated data schema from the policy panel and uses it to analyze the query structure, parsing the query for later stages. The system can thus run a preliminary policy check on the incoming query and route it to the corresponding database proxy. \newline
\noindent \textbf{Original database}: The database service stores the original sensitive data securely to satisfy desiderata \textbf{(III)}, ideally with encrypted storage and restricted access for the data steward and other necessary system security components to fully comply with \textbf{(III)}. \newline
\noindent \textbf{Budget manager}: With the information about both the database schema and the sensitivity annotation from the policy panel, the budget manager models the differential privacy budget and keeps track of the budget consumption in various queries. \newline\newline
\noindent \textbf{Differentially private database proxy}: Before executing the query from the data analyst on the tables in the original database, the proxy analyzes how to apply differential privacy by transforming the query and also calculates the budget consumption to satisfy desiderata \textbf{(I), (II)} and \textbf{(VI)}. Before returning the query result, it also outputs the query's accuracy estimation to satisfy \textbf{(VII)}. \newline
\noindent \textbf{Synthesized database}: The synthesized database maintains the dummy or differential-privately synthesized versions of tables in the original database to satisfy desiderata \textbf{(IV)}. \newline
\noindent \textbf{Synthetic database proxy}: Upon receiving queries to the dummy or differential-privately synthesized data, the proxy checks whether the required version of the tables has already been generated in the synthesized database. If the required version is missing, the proxy orchestrates the generation procedure from the original database on-demand.
\smallskip

Lastly, we describe the communication between system components to explain the workflow to access different privacy-enhancing analytic functionalities.

\noindent \textbf{(1) Data analyst user registration.} Once the system is correctly set up, data analysts should begin to create their user accounts in the system with the user registration service. \newline 
\noindent \textbf{(2) Submission of the query request.} The request should include both the SQL query and a piece of metadata that specify the privacy details like accuracy requirements or whether to use the dummy or synthesized data. The query gateway checks the query request to see if it is compliant with the system policy and routes it to the corresponding database proxy. \newline 
\noindent \textbf{(3) Exploring data statistics on the dashboard.} Data analysts can use the dashboard to explore dataset statistics that are periodically gathered from the original database. \newline
\noindent \textbf{(4) Data steward adjustment for data access policy.} In addition to allowing the data steward to configure the data access policy manually, ideally, the policy panel should also make suggestions on potentially useful policy changes. Such suggestions can be based on the data access application or frequently rejected requests to other system components during user registration. \newline
\noindent \textbf{(5) User registration following policies in the policy panel.} If the data steward decides to include specific steps during the registration procedure (e.g., signing acknowledgments, reading materials, finishing tutorials), the registration procedure would reflect such requirements. \newline
\noindent \textbf{(6) User access to query and dashboard service deter-
mined by policy.} Considering both queries and the statistic dashboard exploration reveal information about the original data, the policy panel should control the access of data analysts to both services.
\noindent \textbf{(7) Query execution with one proxy.} Depending on the privacy-related metadata, one of the proxies executes the query with its transformation and returns the query result with privacy details like result accuracy or budget consumption. If the result consumes the privacy budget, the proxy also notifies the budget manager to track the change. \newline
\noindent \textbf{(8) On-demand data synthesis.} If the synthetic database proxy cannot find the required version of the dummy of synthetic tables from the synthesized database, it triggers the generation of that required version. \newline
\noindent \textbf{(9) Unified schema synchronization between system components.} The database schema component enforces consistency by tracking the changes in the original database. It locks the whole system for schema changes until the updates are applied to all system components. % The policy panel provides sensitivity annotation for the schema.
\newline 

As of September 2022, our GitHub hosts an early-stage open-source effort to benchmark libraries and frameworks suitable for some of the system components in Fig.~\ref{fig:system_design}: \newline

\noindent \centerline{\url{https://github.com/camelop/dp_lab}}

\end{document}